\documentclass[a4paper,11pt]{article}
\usepackage{jinstpub} 
\usepackage{lineno}

\title{\boldmath Design \& Optimization of the HV divider for JUNO 20-inch PMT}

\author[a]{Fengjiao Luo,}

\author[b,c,1]{Zhimin Wang,\note{Corresponding author.}}

\author[b]{Anbo Yang,}

\author[b,c]{Yuekun Heng,}

\author[b,c]{Zhonghua Qin,}

\author[b,c]{Meihang Xu,}

\author[b,c]{Sen Qian,}

\author[b,c]{Shulin Liu,}
\author[b,c]{Yifang Wang,}

\author[d]{Wei Wang,}

\author[e]{Alexander Olshevskiy}

\author[f]{Guorui Huang,}
\author[f]{Zhen Jin,}
\author[f]{Ling Ren,}
\author[f]{Xingchao Wang,}
\author[f]{Shuguang Si,}
\author[f]{Jianning Sun.}
 
 \affiliation[a]{University of South China, School of Nuclear Science and Technology, Hengyang 421001, China}
 \affiliation[b]{Institute of High Energy Physics, Chinese Academy of Sciences, Beijing 100049, China}
 \affiliation[c]{State Key Laboratory of Particle Detection and Electronic, Beijing 100049, China}
 \affiliation[d]{Sino-French Institute of Nuclear Engineering and Technology, Sun Yat-sen University, Zhuhai, 519082, China}
 \affiliation[e]{Joint Institute for Nuclear Research,Dubna, 141980, Russiay}
 \affiliation[f]{Northern Night Vision Technology Co., LTD,211100, Nanjing, China}

\emailAdd{wangzhm@ihep.ac.cn}

\abstract{The Jiangmen Underground Observatory (JUNO) is a 20-kton liquid scintillator detector that employs 20,012 20-inch photomultiplier tubes (PMTs) as photon sensors, with 5,012 dynode-PMTs from HAMAMATSU Photonics K.K.~(HPK), and 15,000 MCP-PMTs from North Night Vision Technology (NNVT) installed in pure water. JUNO aims to provide long-lasting and the best performance operation by utilizing a high-transparency liquid scintillator, high detection efficiency PMTs, and specially designed electronics including water-proof potting for the high voltage (HV) dividers of PMTs. In this paper, we present a summary of the design and optimization of HV dividers for both types of 20-inch PMTs, which includes collection efficiency, charge resolution, HV divider current, pulse shape, and maximum amplitude restriction. We have developed and finalized four schemes of the HV divider for different scenarios, including the final version selected by JUNO. All 20,012 20-inch PMTs have successfully undergone production and burning tests.}


\arxivnumber{2307.10544} 

\begin{document}
\maketitle
\flushbottom

\section{Introductions}
\label{sec:intro}
\begin{figure}[!ht]
    \centering
	\centering
	\includegraphics[width=10cm]{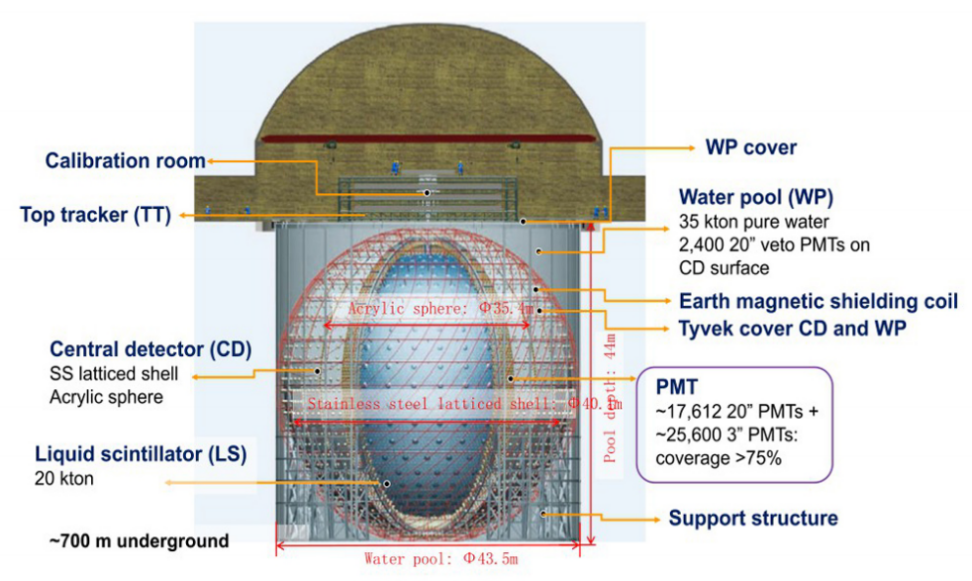}
	\caption{Schematic view of the JUNO detector\,\cite{JUNO:physics:2022PrPNP.12303927J}}
    \label{Figure 1}
\end{figure}

The Jiangmen Underground Neutrino Observatory (JUNO) is a multipurpose neutrino experiment that has been designed to achieve a wide range of objectives. Its primary goal is to determine the ordering of neutrino masses at the 3-4 sigma level in six years of data taking and to measure neutrino oscillation parameters by detecting reactor neutrinos with an accuracy of less than 1\%. JUNO is also capable of observing supernova neutrinos and studying atmospheric, solar, and geo-neutrinos, in addition to performing searches for exotic phenomena\,\cite{ref8}. The JUNO experiment is located in Kaiping, Jiangmen, Guangdong province of China. It is situated about 53\,km away from the Yangjiang and Taishan nuclear power plants, with a planned total thermal power of 36\, GW$_{th}$. 
To further suppress muon-induced backgrounds, the detector will be located deep underground with a 700\,m rock overburden\,\cite{ref9}. Fig.\,\ref{Figure 1} shows the schematic of the JUNO detector.

The detector system of JUNO consists of several sub-detectors, which includes the central detector (CD)\,\cite{CD-ref10}, water pool (WP, a water Cerenkov detector)\,\cite{ref11}, and top tracker (TT)\,\cite{ref12}. The CD is an acrylic vessel with a diameter of 35.4\,m filled with 20-kton 1iquid scintillator, which uses 17,612 20-inch PMTs\,\cite{ref13,ref14,ref15} and 25,600 3-inch PMTs as photon sensors\,\cite{JUNO:3inchPMT:CAO2021165347,ref16,ref17}. The water Cerenkov detector is a cylindrical pool equipped with 2,400 20-inch PMTs filled with 35-kton of ultra-pure water as a muon veto and radioactive shielding. The top tracker detector will use to tag muon tracks precisely. 

To reach the expected sensitivity of neutrino mass ordering, the energy resolution of JUNO has to be better than 3\% at 1\,MeV, which requires a scintillation light collection of at least 1,200\,photoelectrons(pes)/MeV \cite{ref20}. To obtain a good energy resolution, the PMTs need to have high photon detection efficiency (PDE) and good performance. The 20-inch PMT system of JUNO \cite{JUNO:20inchPMT:2022hlz,ref18,ref19} includes 5,000 dynode PMTs R12860-50 from Hamamatsu Photonics K.K.\,(HPK) and 15,012 micro-channel plate PMTs (MCP-PMTs) from North Night Vision Technology Co., Ltd.\,(NNVT). The 3-inch dynode PMTs are from Hainan Zhuanchuang Photonics Technology Co., Ltd.\,(HZC). 
The PMT are commonly used in nuclear physics, particle physics, and astrophysics, such as Super-Kamiokande (Super-K) \cite{ref1}, Daya Bay \cite{ref2-dyb}, Borexino \cite{ref3}, Chooz \cite{ref4}, Double Chooz \cite{ref5}, and Large High Altitude Air Shower Observatory (LHAASO) \cite{ref7}, and so on.
The performances of PMTs are not only related to the processing technology of PMTs themselves but also the design of the High Voltage (HV) divider (HV divider or base). The PMTs of JUNO will be submerged in pure water, whereas the PMTs with the HV divider should be sealed and water-proof, and have good stability and lifetime. Furthermore, the HV divider of the newly developed 20-inch MCP-PMT is one of the key features differentiating it from traditional dynode-PMTs. 

This paper focuses on the R\&D, optimization, production, and quality control of HV dividers for JUNO 20-inch PMTs. The paper is structured as follows: In Sec.\ref{sec:artwork}, the requirements and pre-design of the HV divider for JUNO 20-inch PMTs are briefly introduced. In Sec.\ref{sec:1:opt}, the optimization of the HV divider is discussed, including the collection efficiency of MCP-PMT, charge resolution of MCP-PMT, HV divider current, pulse shape, and restriction on maximum amplitude. In Sec.\ref{sec:1:finalversion}, the final designs of four options, including JUNO selection, is presented. In Sec.\ref{sec:1:product}, the burning test and quality control are described. In Sec.\ref{sec:1:sum}, a summary is provided. 

\section{Requirements \& Pre-design of HV divider for JUNO 20-inch PMTs}
\label{sec:artwork}
\subsection{Requirements}
\begin{figure}[!htbp]
 	\centering
 	\includegraphics[width=8cm]{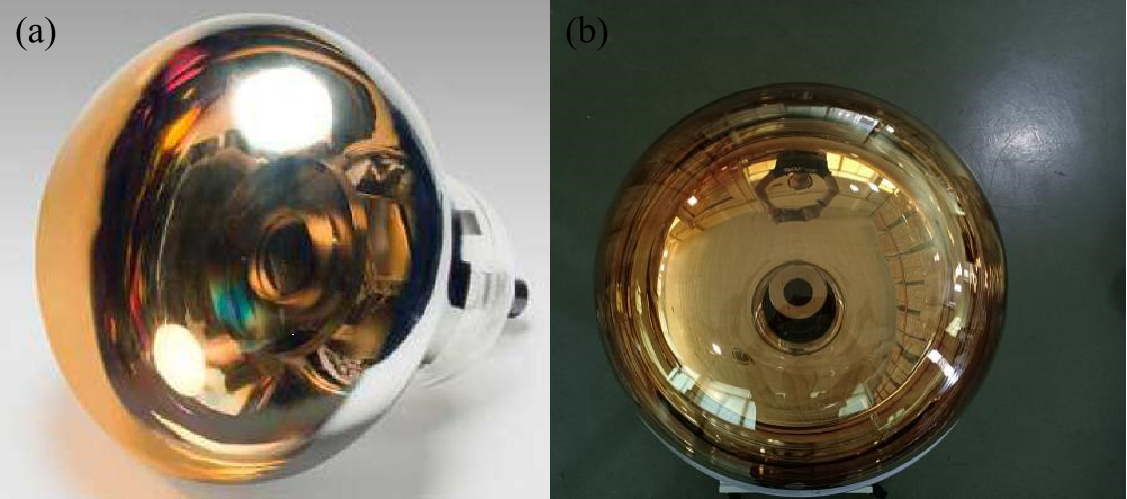}
 	\caption{Physical view of JUNO 20-inch PMTs. (a) 20-inch dynode-PMT; (b) 20-inch MCP-PMT.}
     \label{Figure 2}
 \end{figure}

To achieve high energy resolution, as little light as possible needs to be lost, so 20012 LPMT will be used. Consequently, a list of PMT performance criteria has been required for all the selected PMTs~\cite{PMT-selection-WEN2019162766,JUNO:20inchPMT:2022hlz}. For this purpose, JUNO has selected two types of 20-inch PMTs: the traditional dynode-PMT and the innovative micro-channel plate (MCP) PMT, as depicted in Figure \ref{Figure 2}.
 
 In addition to the performances, the system lifetime is also required for the planned running period, and the HV divider cannot be disassembled once the PMT waterproof is completed. The HV divider of the PMT is a critical component that directly affects its function, performance, and reliability. Therefore, the design of the HV divider is crucial to ensuring the performance and lifetime of JUNO. The JUNO collaboration has established strict requirements for the PMT HV divider, as shown in Table \ref{table 2}.
 \begin{table}
	\begin{center}
		\caption{Key parameters of 20-inch PMT HV divider required by JUNO}
        \label{table 2}
		\begin{tabular}{cc}
			\hline \\
			Parameters &	Requirements \\
			\hline  
			Current of HV divider  & Dynode-PMT: \textless~300\,\textmu A
			\\  &   MCP-PMT: \textless~300\,\textmu A  \\
   			
			Gain &	$10^{7}$ \\
			
			Dynamic Range &	0-4000\,pe \\
			
			Linearity &	1000\,pe@$\pm$10\%  \\
			
			Rise time of waveform &	\textgreater~3\,ns	\\
			
			Overshoot in amplitude & \textless~1\%\\
   			
			Life time &	\textgreater~20 years\\
   			
			Failure rate in 6\, years &	\textless~0.5\% \\
			\hline
		\end{tabular}
	\end{center}
\end{table}

\subsection{Pre-Design}

A high voltage from 500 to 3000\,V is applied usually across the cathode\,(K) and anode\,(A) of a PMT, with a proper voltage gradient among the photoelectron focusing electrode, and dynodes or MCPs. In practice, the inter-stage voltage for each electrode/MCP is provided by voltage-dividing resistors or Zener diodes connected between the anode and cathode\,\cite{ref21}, naming it as an HV divider (or base). For JUNO, a resistor chain is suggested for voltage-dividing taking into account the reliability and simplicity of electronic components.

Table \ref{table 3} shows the suggested voltage ratio of the two types of PMTs from the vendors. Further optimization of the design to meet the requirements of JUNO is needed, including optimizing the voltage ratio, pulse shape, and working direct current (DC) to limit power consumption, and implementing the suggested final design. Additionally, for the newly developed 20" MCP-PMTs, the characteristics of the MCPs needed to be further considered.
\begin{table}
	\begin{center}
		\caption{The HV ratio of 20-inch dynode- and MCP-PMT provided by the vendors.} 
        \label{table 3}
		\resizebox{0.8\textwidth}{!}{
			\begin{tabular}{|c|c|c|c|c|c|c|c|c|c|c|c|c|c|}
				\hline
				dynode-PMT &            K            & Dy1  &  F  & Dy2 & Dy3 & Dy4 & Dy5 & Dy6 & Dy7 & Dy8 & Dy9 & Dy10 & A \\ \hline
				\multicolumn{2}{|c|}{Ratio}       & 11.3 & 0.6 & 3.4 & 
 3.7 &  3  &  2  &  1  &  1  &  1  &  1  &  1  &  1   \\ \hline
				\multicolumn{2}{|c|}{Capacitor (nF)} &      &     &     &     &     &     &     & 10  & 10  & 10  &  10  &   \\ \hline 
			\end{tabular}
        }\\
        \resizebox{0.8\textwidth}{!}{
			\begin{tabular}{|c|c|c|c|c|c|c|c|c|}
				\hline
				MCP-PMT &     K   & F1  &  F2  & MCP1a & MCP1b & MCP2a & MCP2b & A  \\ \hline
				\multicolumn{2}{|c|}{Ratio}       & 7 & 1 & 0 &  10  &  0.68  &  10  &1     \\ \hline
				\multicolumn{2}{|c|}{Capacitor (nF)} &      &     &     &     &     & 10  & 10  \\ \hline
			\end{tabular}
        }
	\end{center}
\end{table}

Positive high voltage is favored in most large-scale neutrino experiments for lower noise, lower cost, and easier installation, such as SNO\cite{SNO-JILLINGS1996421}, Double CHOOZ\cite{ref22,ref23}, etc. PMT with positive high voltage will work with only one readout cable and a de-coupler between HV and signal. However, due to the existence of the de-coupling capacitor, the output pulse of PMT will suffer from an overshoot, which is coupled to the primary pulse, and will generate problems to charge measurements and system triggering, as observed by Double Chooz \cite{ref24}, KamLAND \cite{ref25}, SNO\cite{ref26}, Borexino \cite{ref27} and Daya Bay\cite{ref28}. According to the requirements of JUNO, the overshoot, and small divider-current is part of the key points taken into account in our design for the 20-inch PMTs. We have finished a detailed study on the overshoot of PMT output to a controlled contribution. More details could refer to \cite{ref15,ref29}. 
At the same time, the DC is further required by the power consumption, heating effect in the detector, and reliability.

\begin{figure}[!htbp]
   	\centering
   	\includegraphics[width=7cm]{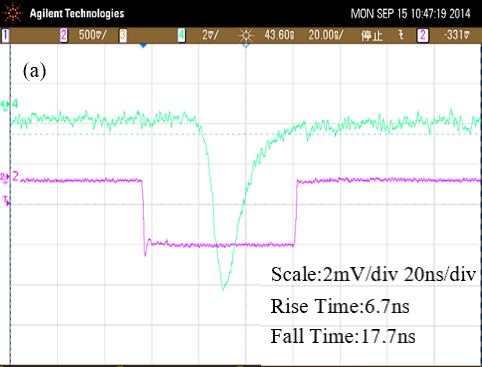}
   	\includegraphics[width=7.1cm]{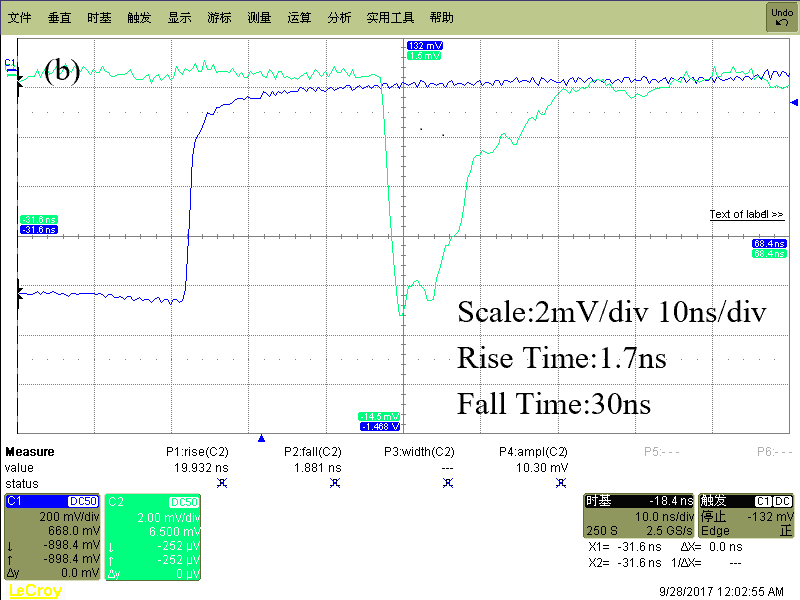}
    \caption{Typical single photoelectron (SPE) waveforms of 20-inch PMTs with pre-designed HV dividers before optimization. (a) Dynode-PMT; (b) MCP-PMT.}
    \label{Figure 5}
   \end{figure}

The typical waveforms of the two types of PMTs can be found in Fig.\,\ref{Figure 5} with the pre-designed HV dividers of the dynode-PMT and MCP-PMT. The rise time and fall time of MCP-PMT do not meet the JUNO requirements. The MCP-PMT waveforms exhibit two peaks and an uneven falling edge, which can impact charge measurement and waveform de-convolution. In the subsequent sections, we will describe our optimization of the HV divider to meet JUNO's requirements for the relevant features.

\section{Optimization}
\label{sec:1:opt}
\subsection{Collection efficiency of MCP-PMT}
The high photon detection efficiency (PDE) of the PMT is a critical requirement for JUNO. PDE is determined by photon transmission, quantum efficiency (QE), and collection efficiency (CE) of the PMT. The collection efficiency is influenced by the shape of the photocathode, the design of the focusing electrode, and the electric field distribution. Without an appropriate applied voltage between the cathode and the first multiplier, photoelectrons (pe) cannot be efficiently collected or amplified, thus affecting the collection efficiency and charge amplification process. In this section, we will focus on optimizing the first several stages of the JUNO 20" MCP-PMT.

\begin{figure}
	\centering
	\includegraphics[width=12cm]{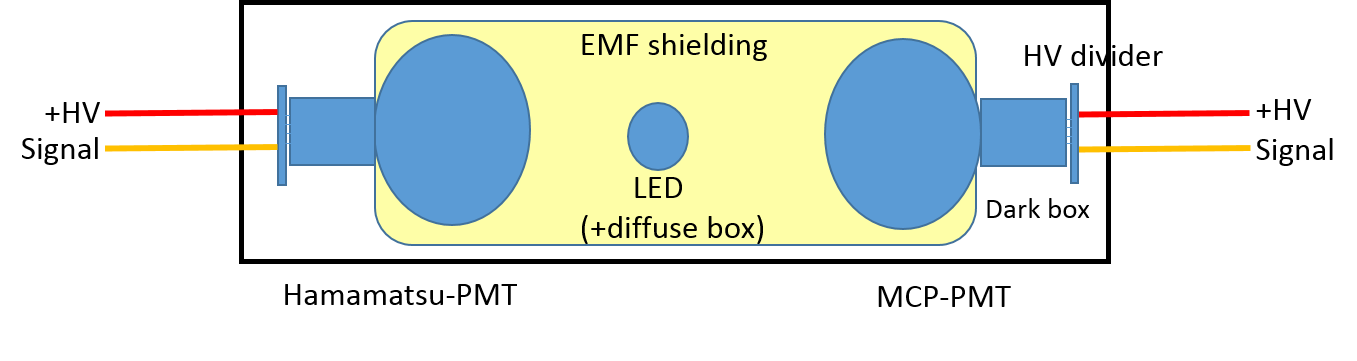}
	\caption{Diagram of the measurement system for the relative detection efficiency. A light source with diffuser ball is located between the two 20" PMTs.}
     \label{Figure 6}
\end{figure}

For an MCP-PMT, the applied voltages between the photocathode, the focusing stage, and the first MCP will affect the internal electric field distribution and the process of photoelectron collection or amplification. Efficient collection of photoelectrons by the first MCP results in a good collection efficiency for the PMT. However, the collection efficiency of PMT cannot be measured directly. A measurement system of relative detection efficiency is set up, as shown in Fig.\,\ref{Figure 6}\cite{relative-eff-anbo}, where a 20-inch dynode PMT is used as a reference. The collection efficiency of photoelectrons will saturate when the collection field is strong enough \cite{haiqiong-relative-collection-efficiency}. The following tests have satisfied this condition.

\begin{figure}[!htbp]
	\centering
	\includegraphics[width=10cm]{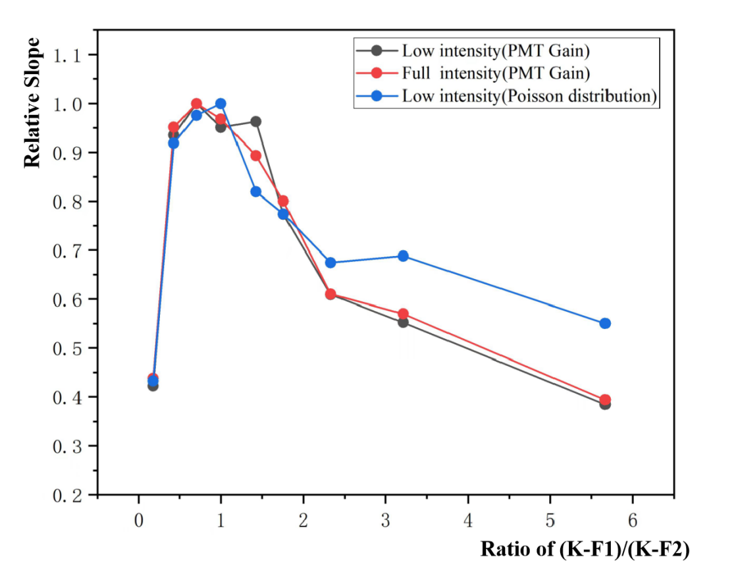}
	\caption{Relative DE relative to the reference PMT from different ranges of light intensity and a survey on different voltage ratios of the K-F1 and F1-F2 for the MCP-PMT. It is calculated from the relative output charge of the MCP-PMT and the reference dynode-PMT in pe, which is related to the solid angle, PDE of each individual PMT. The y-axis is a further normalized value of each curve with the ratio of  (K-F1)/(F1-F2)=1.}
 \label{Figure 7}
\end{figure}

The ratio of K-F1 (cathode to focus stage 1) and F1-F2 (focus stage 1 to stage 2) (in Table \ref{table 3}) has an impact on the collection efficiency of the MCP-PMT. To investigate this, we surveyed the voltage ratio between K-F1 and F1-F2 with the system, applying a strong enough field to collect all photoelectrons. The relative detection efficiency of the MCP-PMT with different ratios (K-F1)/(F1-F2) from 0.2 to 5.0 was measured, as shown in Fig.\ref{Figure 7}. The total applied voltage between the cathode (K) and focus stage 2 (F2) was kept constant. The measured charge of a PMT in pe was calculated based on two methods: Poisson distribution (mu=-ln(1-N\_sig/N\_total))\cite{JUNO-PMT-gain-Zhang2021GainAC,JUNO:20inchPMT:2022hlz} and PMT gain (pe=(S-P)/($e$×Gain)), where S and P (in pC) are the peak locations of the pedestal and signal on the charge spectrum, respectively, of an electron of the MCP-PMT and the reference PMT, and e is the charge of an electron. The output charge relationship between the MCP-PMT and the reference PMT, including different light intensities, was fitted by a linear curve  \cite{relative-eff-anbo} for each ratio of (K-F1)/(F1-F2), and the results of different ratios were compared. It is important to note that the y-axis in 
Fig.\ref{Figure 7} is a normalized value relative to the measurement of the ratio of (K-F1)/(F1-F2) = 1 of each curve itself.

The design of MCP-PMT with respect to the electric field between K-F1 and F1-F2 is crucial. If the electric field of K-F1 or F1-F2 is too high, the photoelectrons will be easily collected and absorbed by F1 or F2 itself, rather than the first MCP for further amplification. This will result in a reduced effective collection efficiency. Only when the ratio between K-F1 and F1-F2 is around one, can the photoelectrons be effectively collected by the first MCP for the following multiplication. With this setting, the field distribution between K, F1, and F2 is optimal, and the photoelectrons can be better collected by the first MCP. For the JUNO MCP-PMT, a voltage ratio of K-F1 to F1-F2 of one is suggested.

\subsection{Charge resolution of MCP-PMT}
To achieve better amplification and charge resolution, MCP-PMTs use two internal MCPs in parallel\,\cite{ref30}. It has been confirmed that the charge resolution of the MCP-PMT output can be improved with an inverted voltage in the gap between the two MCPs (the gap stage MCP1b-MCP2a as shown Fig.\ref{Figure 8} between MCPs)\,\cite{ref31,ref32,ref33}. The inverted voltage reduces the spread of electrons arriving at the second MCP by rejecting those with lower kinetic energies in the periphery and only allowing electrons with higher kinetic energies to pass to MCP2. This feature affects both the amplification gain and the charge resolution of the MCP-PMT. 

\begin{figure}
    \includegraphics[width=8cm]{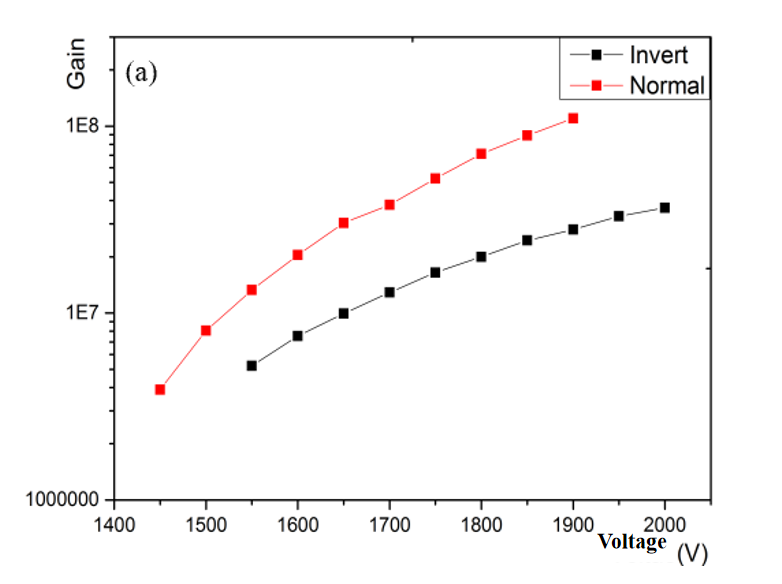}
    \includegraphics[width=7.3cm]{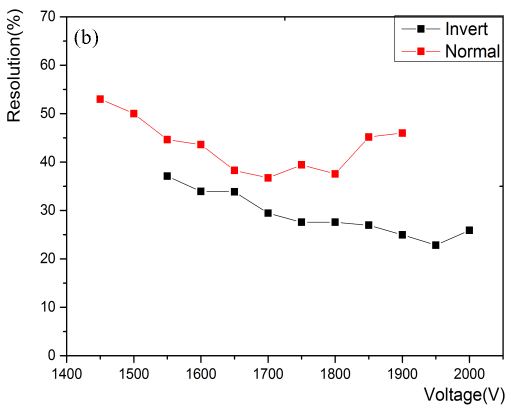}
    \includegraphics[width=9cm]{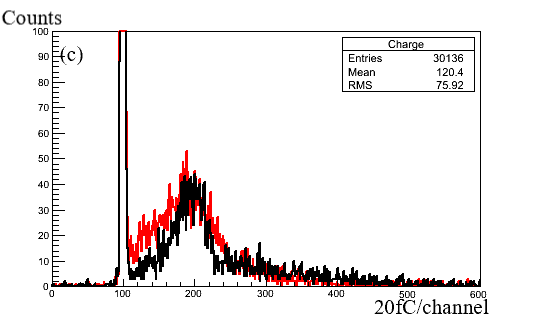}
	\caption{(a) Gain versus voltage of MCP-PMT (serial No.\,16105) with internal normal and inverted gap voltage (black: inverted gap voltage; red: normal gap voltage). (b) Charge resolution (calculated from a Gaussion fitting to the SPE peak) versus voltage of MCP-PMT (serial No.\,16105) with internal normal and inverted gap voltage; (c) SPE spectrum of MCP-PMT with internal normal and inverted gap voltage with a same peak gain\cite{JUNO-PMT-gain-Zhang2021GainAC}.}
    \label{Figure 9}
\end{figure}

\begin{figure}[!htbp]
	\centering
	\includegraphics[width=7cm]{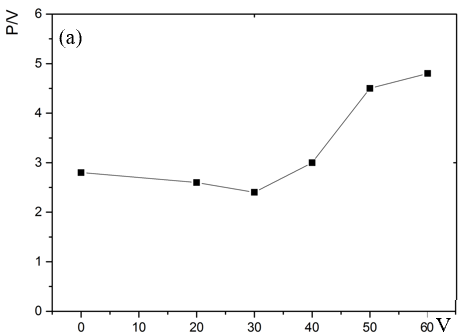}
     \includegraphics[width=7cm]{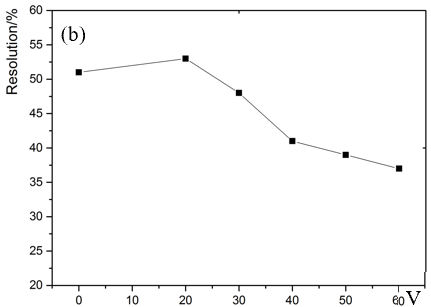}
	\caption{(a) Peak-to-Valley ratio versus inverted gap voltage; (b) Charge resolution versus inverted gap voltage. }
     \label{Figure 10}
\end{figure}

\begin{figure}[!htbp]
	\centering
	\includegraphics[width=14cm]{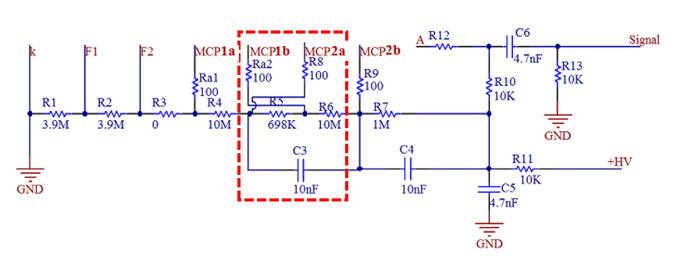}
	\caption{Schema of the HV divider of MCP-PMT with inverted gap voltage between MCPs}
    \label{Figure 8}
\end{figure}

We conducted an evaluation of the effects of the inverted gap voltage on the PMT gain and charge resolution of several PMT samples, using a Gaussian fitting to the spectra SPE(Single Photoelectron) peak. The SPE spectra of the MCP-PMT with both the inverted and normal gap voltage are shown in Fig.\ref{Figure 9}. As expected, the PMT gain reduced from the normal gap voltage to the inverted gap voltage due to fewer electrons reaching the second MCP. A higher voltage was required in total to achieve the same gain for the configuration of the inverted gap voltage. However, the charge resolution of the MCP-PMT with an inverted gap voltage is better than that with the normal gap voltage for the same applied voltage or the same gain, which is consistent with our expectations. It is important to note that the higher the working voltage of a PMT, the larger the dark noise rate. Further checks were conducted with different inverted gap voltages of MCP-PMT in detail on the gain, dark noise, charge resolution of SPE, and peak-to-valley ratio of MCP-PMT, as shown in Table \ref{table 4} and Fig.\ref{Figure 10}. Based on our findings, an inverted gap voltage between the two MCPs of the JUNO MCP-PMT HV divider is suggested, as shown in Fig.\ref{Figure 8}. This voltage should be higher than 30\,V, which can be satisfied by most of the working voltage of MCP-PMTs with a gain of 1$\times 10^7$.

\begin{table}
	\begin{center}
 \caption{Gain and dark noise rate of the MCP-PMT with different inverted gap voltage.}
 \label{table 4}
 \resizebox{0.6\textwidth}{!}{
 \begin{tabular}{cccc}
  \hline
  Gap Voltage & Voltage Value & HV  & Dark Rate \\ 
   &   &  (gain@$10^7$) & (gain@$10^7$)  \\ 
  \hline 
  Inverted &  20\,V & 1560\,V & 15\,kHz   \\
  & 30\,V&	1560\,V & 14\,kHz\\
   & 40\,V &1650\,V & 17\,kHz\\
   & 50\,V	& 1675V & \textless20kHz\\
  & 60\,V &1675\,V&	17\,kHz\\
  Normal & 36\,V & 1535\,V & 14\,kHz\\
  \hline
  \end{tabular}
  }
  \end{center}
\end{table}

\subsection{HV divider direct current(HV DC)}
For JUNO, the high voltage for each PMT will be provided by a high voltage unit (HVU) produced by the HVSYS company\cite{HVU-ru}. However, the maximum current of the designed HVU is less than 300\,\textmu A, which will limit the working DC of the PMT or even further lower it, taking into account the reliability. It is important to note that the higher the working current, the lower the reliability.

For a dynode-PMT, the resistance between the electrodes is virtually infinite, so the resistance of the HV divider chain is almost decoupled from the PMT itself within a large range, allowing for a constant voltage division ratio\cite{ref21}. Based on our analysis, we conclude that the working DC of the HV divider should be less than 1/3 of the maximum DC of the HVU, taking into consideration the reliability, linearity\cite{diru-PMT-linearity}, and component selection.

For an MCP-PMT, the finite body resistance of the MCP is an important feature that distinguishes it from the dynode-PMT. The body resistor of MCP and the HV divider resistor chain will influence each other, which will further affect the performance of the MCP-PMT. The effective resistance ($R$) of the MCP with a designed HV divider can be calculated using Formula \ref{3.1}.

\begin{equation}\label{3.1}\tag{3-1}
		R =	\frac{R_{divider}  \times R_{body}}{R_{divider}  + R_{body}},
	\end{equation}

The $R_{divider}$ represents the resistance of R4 and R6 in Fig.\ref{Figure 8}, which is in parallel with the MCPs, while the $R_{body}$ represents the body resistance of the MCP. When the $R_{divider}$ is much smaller than the $R_{body}$, the effective resistance $R$ is approximately equal to the $R_{divider}$, and the divider ratio of the HV divider will not be influenced by the $R_{body}$. The mean value of $R_{body}$ is around 20\,M$\Omega$ for the 15,000 JUNO MCP-PMTs. Taking into consideration the possible options of the HV divider DC of around 180\,\textmu A (High DC) and 70\,\textmu A (Low DC), we compared the peak-to-valley ratio and charge resolution, as shown in Fig.\ref{Figure 9} and Fig.\ref{Figure 10}. The charge distribution of SPE of high DC is slightly narrower than that of low DC.

We also measured the non-linearity of the MCP-PMT with different DCs, as shown in Fig.\ref{Figure 9}. When the intensity is small (at a single photoelectron level), the measured non-linearity of PMT response is deeply affected by the system noise and statistical fluctuations. When the intensity is less than 1000\,pe (with a frequency of 200\,Hz), there is no obvious difference in non-linearity between the DCs. The peak-to-valley ratio of MCP-PMT and the charge resolution with High DC are better than that of Low DC. When the MCP-PMT is working at the same gain, the working voltage at high DC is lower than that of low DC. However, there is no obvious effect on timing performance among different DCs.

In summary, when the MCP-PMT is working at high DC (180\,\textmu A@1800\,V), its peak-to-valley ratio and charge resolution are better than those of low DC, and the required voltage to achieve the same gain is lower. Based on the requirements for the performance of MCP-PMT, we advise a divider DC of 180\,\textmu A, which has been adopted by JUNO for a balance of performance and acceptable reliability.

\begin{figure}[!htbp]
	\centering
	\includegraphics[width=12cm]{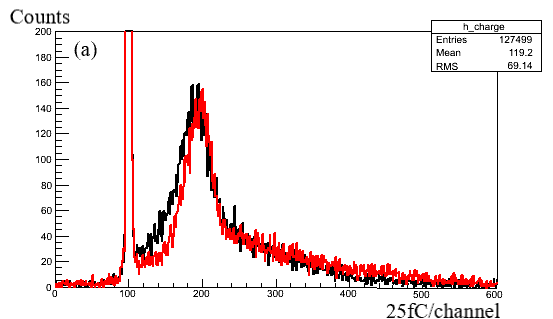}
 \includegraphics[width=11cm]{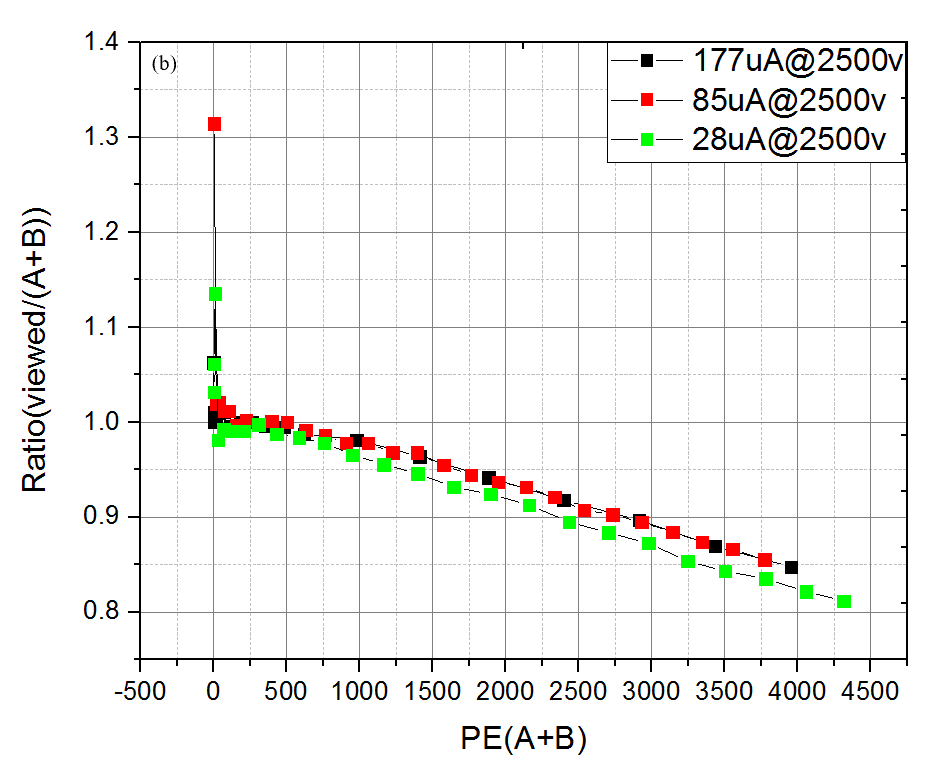}
	\caption{(a) The SPE spectrum of MCP-PMT with divider current of 70\,\textmu A and 180\,\textmu A; (b) The non-linearity of MCP-PMT with different DC.}
    \label{Figure 11}
\end{figure}

\begin{figure}[!htbp]
	\centering
	\includegraphics[width=7.5cm]{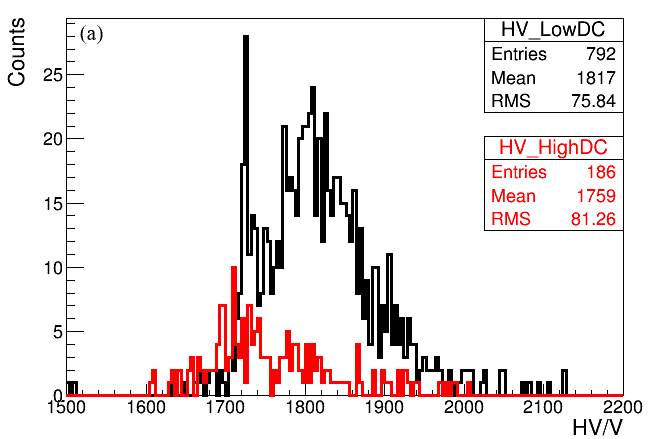}
 \includegraphics[width=7.5cm]{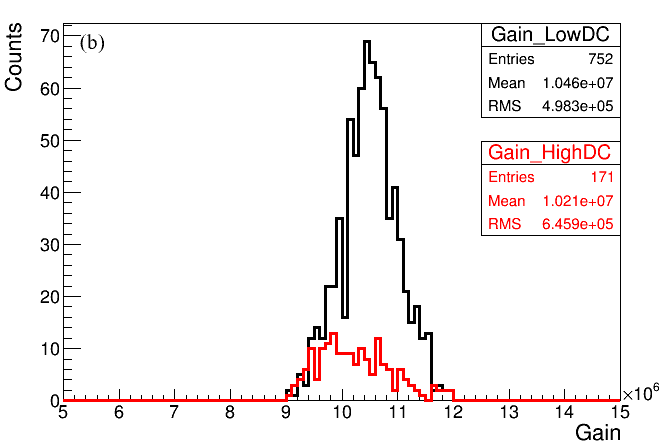}
 \includegraphics[width=7.5cm]{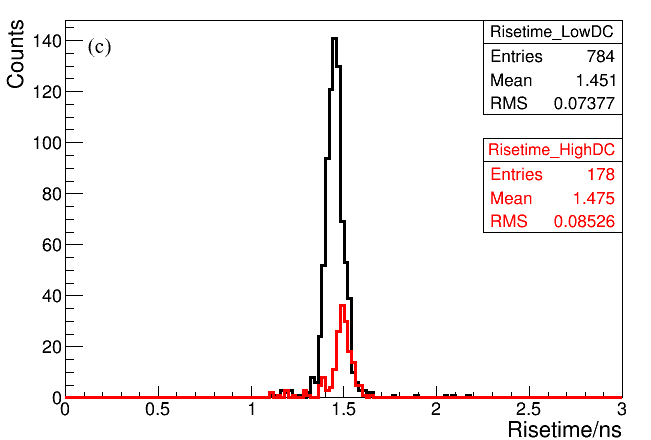}
 \includegraphics[width=7.5cm]{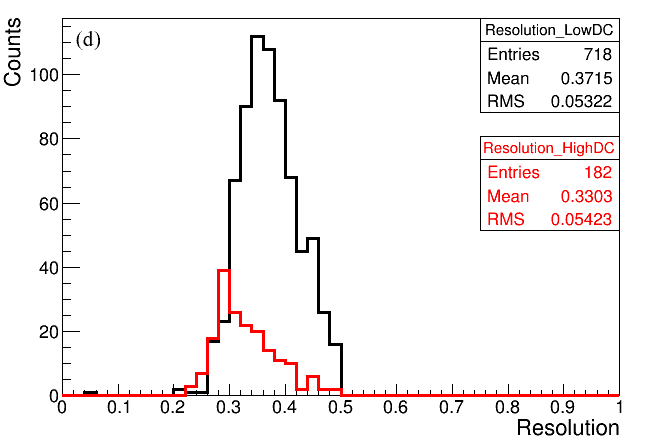}
 \includegraphics[width=7.5cm]{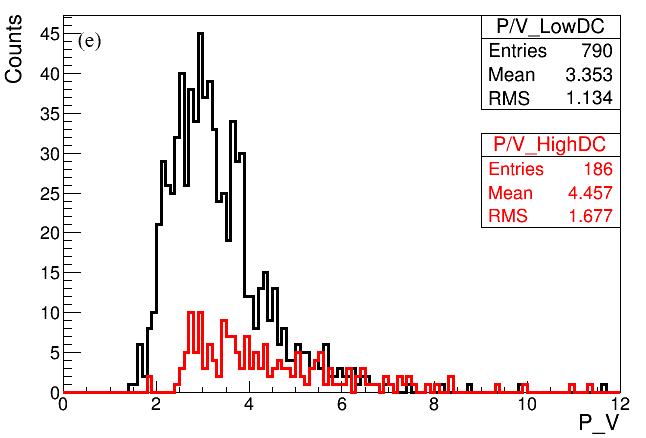}
 \includegraphics[width=7.5cm]{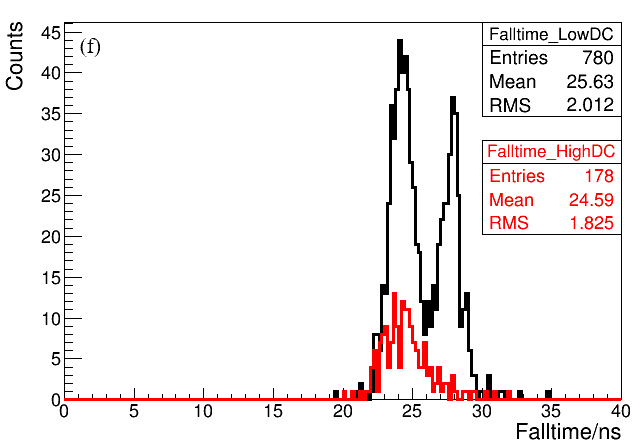}
\caption{(a) The working voltage distribution of the MCP-PMT at a gain of $10^7$; (b) The gain distribution of MCP-PMT with the same voltage; (c) Peak-to-valley ratio; (d) Energy resolution; (e) Rise time; (f) Fall time. }
    \label{Figure 12}
\end{figure}

\subsection{Pulse shape}

\begin{figure}[!htbp]
	\centering
	\includegraphics[width=15cm]{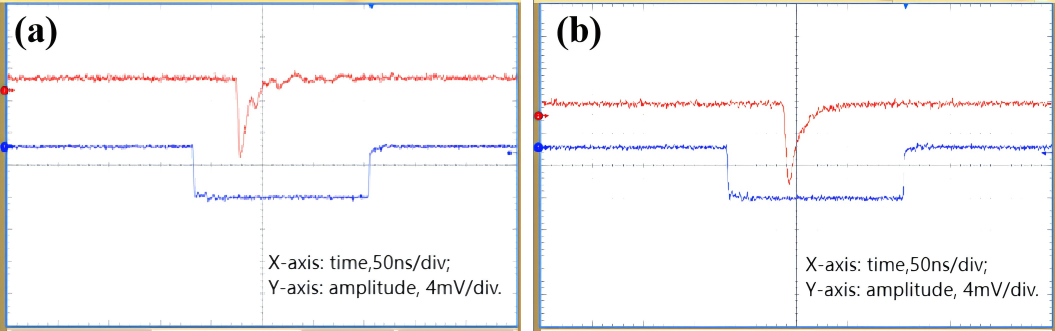}
	\caption{The SPE waveforms of MCP-PMT with different configurations for ringing.
	(a) Ra1= Ra2= R8= R9= 0\,$\Omega$;
	(b) Ra1= Ra2= R8= R9= 50\,$\Omega$;
        }
        
     \label{Figure 14}
\end{figure}

\begin{figure}[!htbp]
	\includegraphics[width=7.5cm]{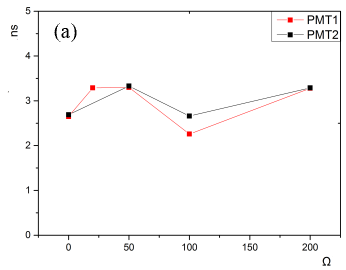}
    \includegraphics[width=7.5cm]{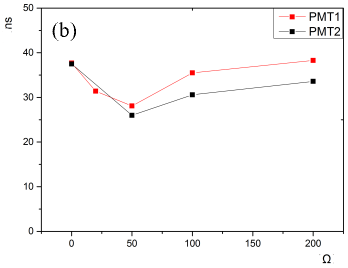}
    \includegraphics[width=7.5cm]{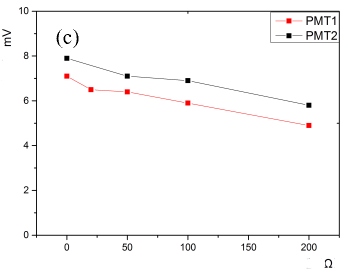}
	\caption{(a) Rise time of SPE versus resistors. (b) Fall time of SPE versus resistors. (c) Amplitude of SPE versus resistors.}
   \label{Figure 15}
\end{figure}

In a large-scale high-precision neutrino experiment, the quality of the PMT output pulse is one of the key factors in achieving accurate measurements. In the circuit shown in Fig.\,\ref{Figure 8}, resistors Ra1, Ra2, R8, R9, and R12 effectively reduce the ringing of the output waveform by forming an internal filter. Fig.\,\ref{Figure 14} shows the waveforms of the MCP-PMT with different configurations, and the ringing is efficiently reduced when the resistors Ra1=Ra2=R8=R9=50\,$\Omega$. In Fig.\,\ref{Figure 14}, the red line represents the output waveform of a single photoelectron, and the blue square waveform is the trigger gate synchronized to the signal. The resistors will also affect the rise time, fall time, and amplitude of the SPE. The performances of MCP-PMT are measured as shown in Fig.\,\ref{Figure 15}c. The larger the resistance, the smaller the amplitude of the output waveform. Finally, we suggest using Ra1=Ra2=R8=R9=50\,$\Omega$ for the JUNO MCP-PMT.

\begin{figure}[!htb]
	\centering
	\includegraphics[width=6cm]{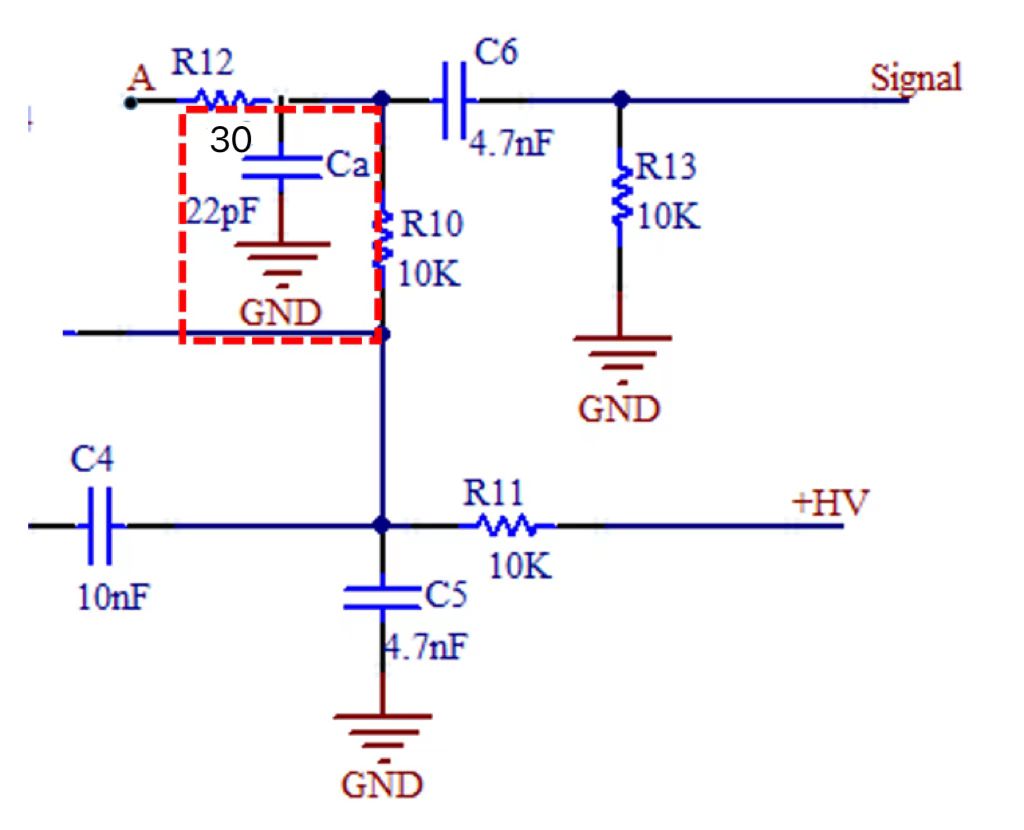}
	\caption{Optimized output circuit for MCP-PMT}
    \label{Figure 16}
\end{figure}

In JUNO, a Flash ADC with 1\,GSample/s is used to sample the waveform of the PMT, and the rise time of the raw waveform of MCP-PMT is about 1-2\,ns, which is determined by the structure of MCP-PMT. The distance between the MCPs and the thickness of the MCP itself are very small order of few millimeter, and the amplification process of electrons is very quick\cite{ref34}. This is much faster compared to the rise time (5\,ns) of Hamamatsu dynode-PMT and the bandwidth of the sampling electronics. As a result, only 1-2 points can be sampled on the rising edge of the pulse, which can lead to information loss. Therefore, it is prescribed to slow down the rise time of MCP-PMT for JUNO. A specialized filter circuit (a capacitor, Ca) for MCP-PMT is designed to slow down its rise time, as shown in Fig.\ref{Figure 16}, where Ca and R12 are composed of an RC filter (\textit{$\tau$} is around 2\,ns, the high-frequency part is filtered out by the RC circuit), 
where the exact value will be confirmed in upcoming measurements because of distributed factors. Without the capacitor of Ca, the waveform of SPE will always have two peaks, as seen in Fig.\ref{Figure 5}, which is due to the impedance mismatching among the internal structure of MCP-PMT. With the Ca, the waveform is smooth and the rising edge is slowed down, and the fall time of MCP-PMT is also affected to around 25\,ns, which is related to the value of R12.

\begin{figure}[!htbp]
	\includegraphics[width=8cm]{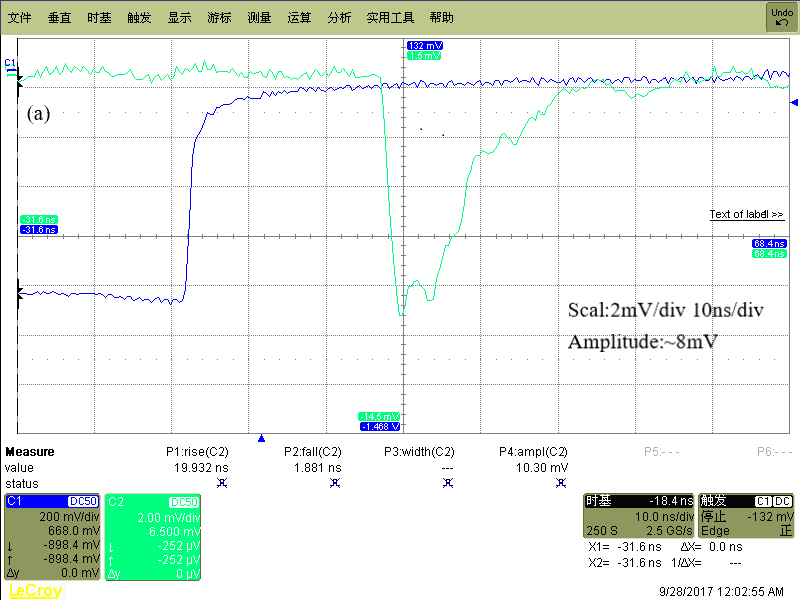}
 \includegraphics[width=8cm]{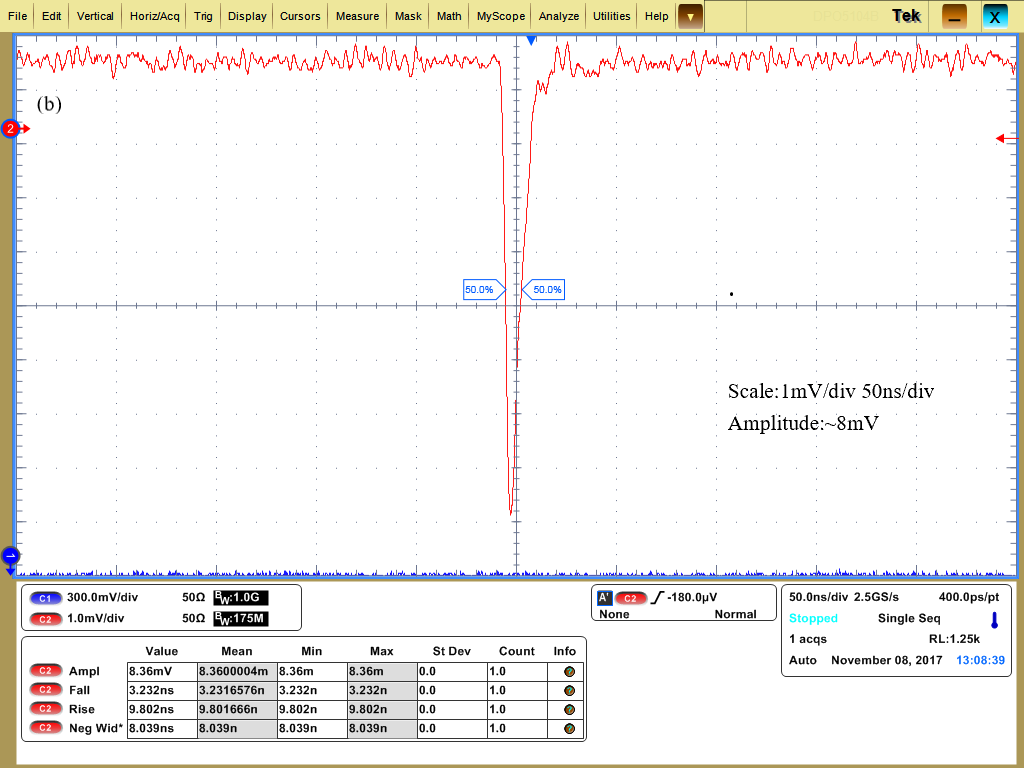}
 \includegraphics[width=9cm]{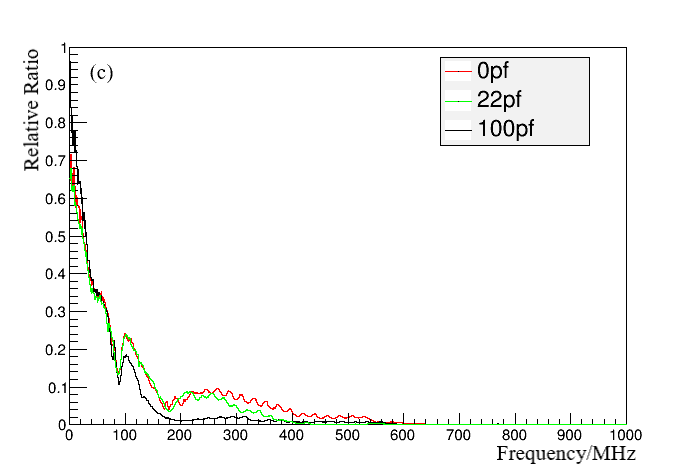}
	\caption{Waveform of SPE at different Ca. (a) Ca=0\,pF; (b) Ca=22\,pF; (c) Spectrum of waveform of SPE under different Ca in frequency domain: the x-axis is the frequency of the spectrum in MHz, and the y-axis is the power.}
    \label{Figure 17}
\end{figure}

We surveyed the value of R12 while keeping the other values constant. The value of R12 affects the fall time, amplitude, and ringing of the waveform by forming another RC circuit. The smaller the R12, the faster the fall time and the greater the amplitude. Additionally, the ringing of the waveform is related to the R12, with smaller R12 values leading to larger ringing.When R12 is equal to 20\,$\Omega$, the ringing is acceptable for JUNO, as shown in Fig.\ref{Figure 17}. However, to further reduce the ringing, we advise using a value of R12 of 30\,$\Omega$. After optimizing the rise time of the waveform, the fall time is also changed.

The features of MCP-PMT with different configurations are measured and summarized in Table \ref{table 5}. The rise time of the waveform can be slowed down to less than 3\,ns when the Ca is set to 22\,pF or 30\,pF, where the high-frequency part of the waveform is filtered by the RC circuit. Fig.\ref{Figure 17} shows the waveform at Ca=0\,pF and Ca=22\,pF when keeping the R12 unchanged. Fig.\ref{Figure 17} also shows the waveform of SPE after applying the Fourier transforms with different Ca values. From the spectrum of the waveform in the frequency domain, we observe that the greater the Ca, the larger the waveform distortion.

In summary, based on the rise time, the shape of the waveform, and the optimal filtering of high-frequency components, we suggest using Ca=22\,pF and R12=30\,$\Omega$ as the optimal choice for JUNO MCP-PMT.

\begin{table}  	
	\begin{center}
		\caption{The parameters of MCP-PMT with different Ca}\label{table 5}
		\resizebox{1\textwidth}{!}{
			\begin{tabular}{cccccccccc}
				\hline
				&  Ca  &  Gain  & Voltage & PV & Charge & Rise & Fall & FWHM & SPE \\
				&  pF  & $10^7$ &    V    &       ratio     &    Res.  \%         &    time/ns     &      time/ns      &  ns   & Amp./mV   \\ \hline
				PMT1 &  0   &  1.17  &  1770   &         6.7          &        21         &    1.9    &     28.5     & 10.9  &  6.7  \\
				&  10  &  1.17  &  1770   &         7.0          &        22         &    2.5    &     28.9     & 11.0  &  6.7  \\
				&  22  &  1.17  &  1770   &         6.3          &        23         &    3.3    &     28.1     & 12.2  &  6.4  \\
				&  30  &  1.17  &  1770   &         7.4          &        21         &    3.3    &     29.8     & 12.1  &  6.3  \\
				PMT2 &  10  &  1.17  &  1705   &         4.6          &        32         &    2.5    &     27.3     & 10.6  &  7.2  \\
				& 22nF &  1.17  &  1705   &         4.9          &        34         &    3.3    &     26.0     & 10.6  &  7.0  \\
				&  30  &  1.17  &  1705   &         4.3          &        31         &    3.3    &     26.4     & 11.1  &  6.8  \\ \hline
		\end{tabular} }
	\end{center}
\end{table}

\subsection{Restrict the maximum amplitude}

The JUNO detector was established with a rock overburden of 700 meters underground to reduce the backgrounds related to cosmic rays. According to simulations of the actual terrain, there will be around 3\,Hz muons passing through the central detector, and the average energy of muons is 215\,GeV. Muons reaching the CD will be detected and seen by the PMTs. The output strength of high-energy muons passing through the CD will be very large, with thousands to tens of thousands of photoelectrons generated in each PMT. The amplitude of the PMT pulse can reach tens of volts \cite{ref6} if there is no saturation or limitation. For a JUNO MCP-PMT, the maximum output amplitude can reach around 30 to 50\,V\cite{diru-PMT-linearity}, while the maximum output amplitude of HPK PMT is around 7-8\,V with a load of 50\,$\Omega$. The designed dynamic range of JUNO electronics is from 0-8\,V with a load of 50\,$\Omega$. Working under such a large dynamic range to cover from a few millivolts to several tens of volts with high accuracy is challenging and potentially dangerous for the electronics system.

\begin{figure}[!htbp]
	\centering
	\includegraphics[width=6cm]{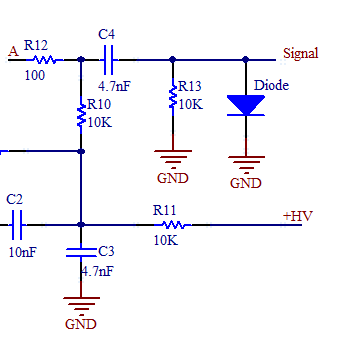}
	\caption{Schematic diagram of MCP-PMT protection circuit}
     \label{Figure 21}
\end{figure}

To limit the maximum output amplitude of the MCP-PMT and meet JUNO requirements, we considered a limitation circuit to restrict the largest amplitude while maintaining an acceptable waveform frequency and reliability. In the circuit design, a diode is suggested as a protection and restriction device, as shown in Fig.\ref{Figure 21}.

\begin{figure}[!htbp]
	\centering
	\includegraphics[width=7.0cm]{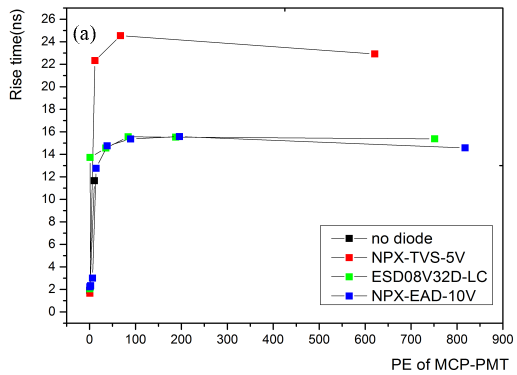}
    \includegraphics[width=7.0cm]{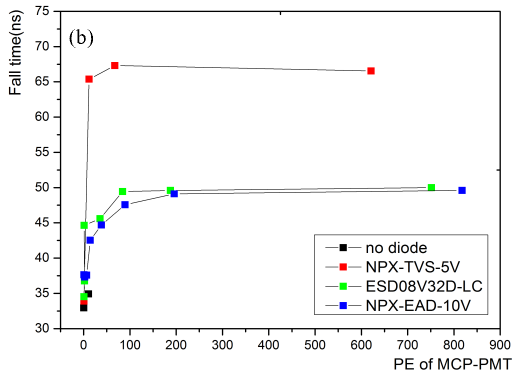}
   \includegraphics[width=7.0cm]{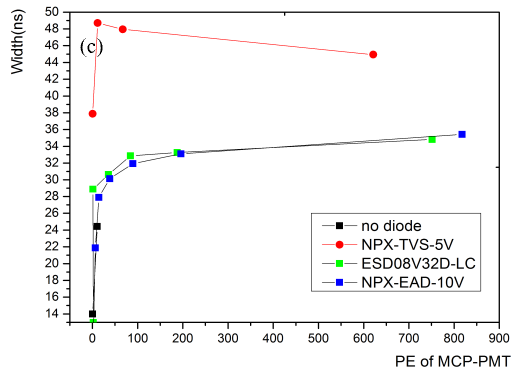}
    \includegraphics[width=7.0cm]{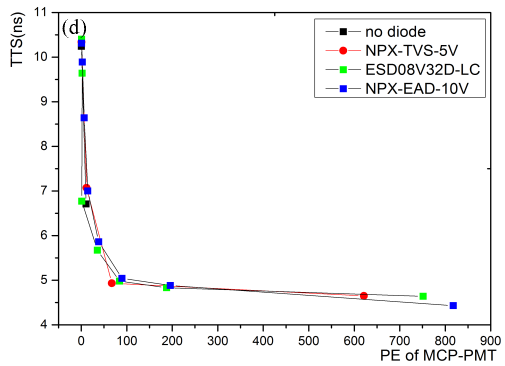}
	\caption{Timing performance of MCP-PMT versus the incident photoelectrons of MCP-PMT with different diodes. (a) Rise Time; (b) Fall Time; (c) FWHM; (d) TTS. }\label{Figure 22}
\end{figure}

Table \ref{table 6} shows the output amplitude of several tested diodes for the MCP-PMT. Without a protective diode, the maximum output of the MCP-PMT is around 28\,V (200\,Hz). With diode protection, the maximum output of the MCP-PMT is limited, which is related to the cut-off voltage of the tested diode. In Table \ref{table 6}, the amplitude of the PMT output begins to decrease when the PMT output voltage exceeds the cut-off voltage of the diode. According to the frequency domain of the waveform, the bandwidth of the PMT waveform is less than 500\,MHz. According to the RF interface parameters, as shown in Table \ref{table 7}, when the diode capacitance is less than 3.5\,pF, there is no effect on the signal. Figure \ref{Figure 22} shows the timing parameters of SPE under four different diodes in Table \ref{table 6}.

\begin{table}  	
	\begin{center}
		\caption{MCP-PMT output amplitude with different diodes}\label{table 6}
		\resizebox{0.8\textwidth}{!}{
			\begin{tabular}{cccc}
				\hline
				Type     & Max output & Specification  & Capacitors (pF) \\
                  &  amp. (V) & Clam voltage (V) & typical (max.) \\ 
  				\hline  
				No diode   &        ~28        &               -               &             -              \\
				NXP-ESD-10V  &        ~17        &              ~10              &        1.55 (1.75)         \\
				NXP-TVS-5v  &       ~7.5        &              ~5               &         152 (200)          \\
				ESD08V32D-LC &        ~15        &              ~8               &            1.2 \\
				\hline
		\end{tabular} }
	\end{center}
\end{table}

\begin{table}  	
	\begin{center}
		\caption{RF interface requirements (Reference criterion: 50\,$\Omega$ system, port voltage standing wave ratio degraded from 1 to 5)}\label{table 7}
		\resizebox{0.8\textwidth}{!}{
			\begin{tabular}{cccc}
				\hline
				Frequency      & Cp  & Frequency & Cp  \\
                 &  (Appro. value) & & (Appro. value) \\
              \hline
				100\,MHz   &     $\leq{16\,pF} $     &  1.8\,GHz   & $\leq{0.9\,pF} $ (1.75) \\
				450\,MHz   &     $\leq{3.5\,pF}$     &  2.1\,GHz   &  $\leq{0.9\,pF}$ (200)  \\
				800\,MHz    &     $\leq{2\,pF} $      &   3\,GHz    &     $\leq{0.5\,pF}$     \\
				1\,GHz  &     $\leq{1.5\,pF}$     &   5\,GHz    &        $\leq{0.33\,pF} $       \\ \hline
		\end{tabular} }
	\end{center}
\end{table}

When using the NXP-ESD-10V or ESD08V32D-LC diode models, the time features of the MCP-PMT waveform are not affected because the capacitance of these two diodes is small, less than 3.5\,pF. Therefore, the NXP-ESD-10V and ESD08V32D-LC diode models are the optimal choices for limiting the maximum output signal amplitude of the MCP-PMT while maintaining the time features of the waveform.

\section{Final designs}
\label{sec:1:finalversion}

Several versions of the HV dividers have been designed to match the requirements of different options of JUNO electronics for R$\&$D and testing purposes.
\subsection{Design I: w/ restriction on Maximum amplitude}
\begin{figure}[!ht]
	\centering
	\includegraphics[width=12.0cm]{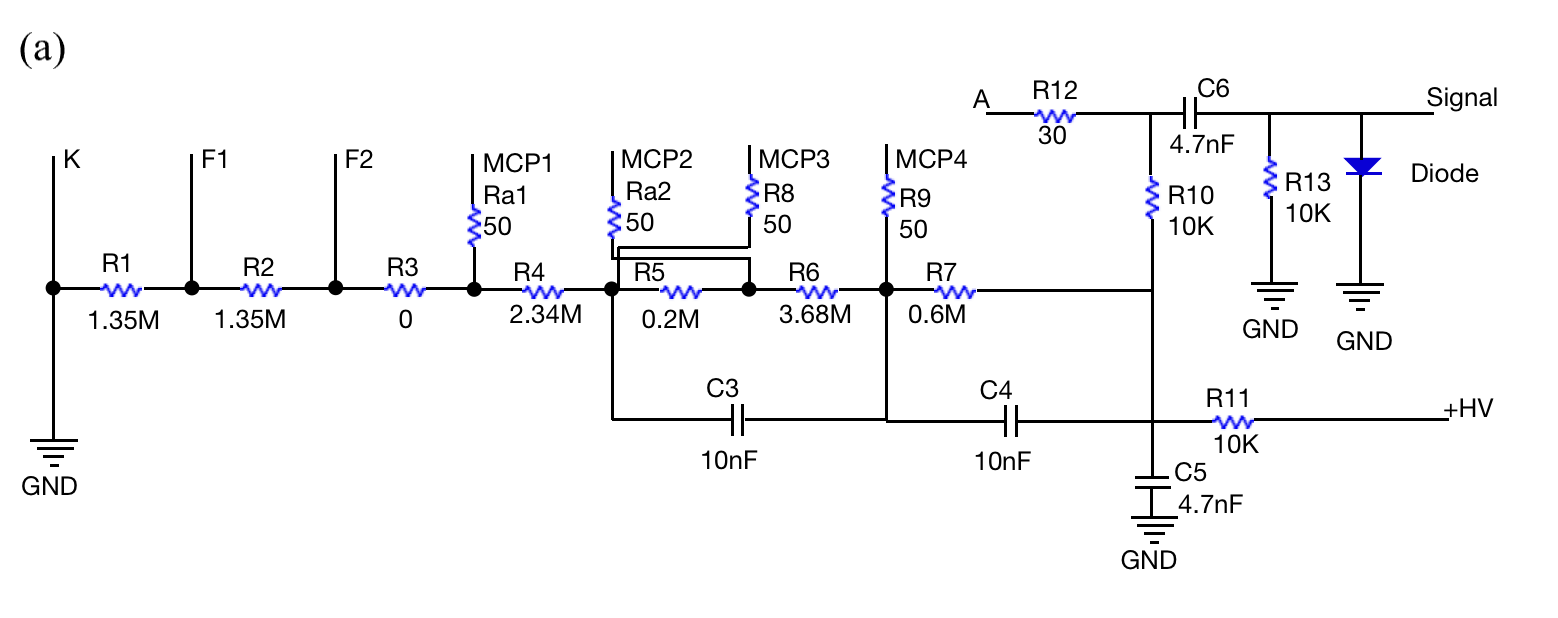}
    \includegraphics[width=12.0cm]{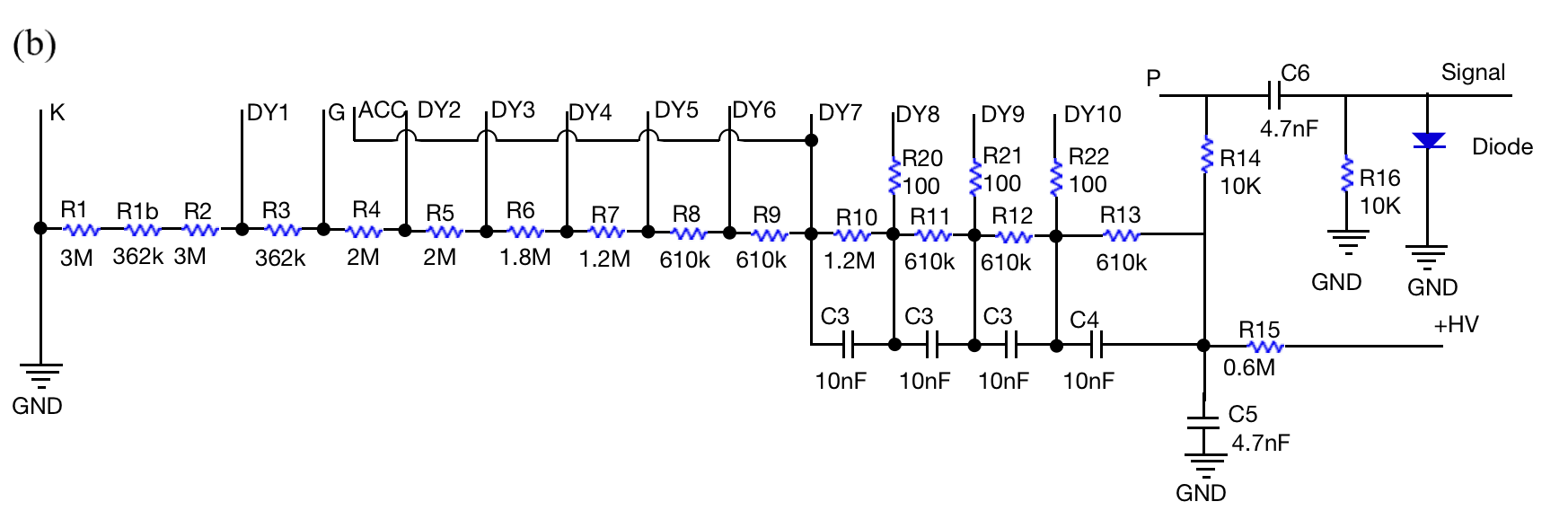}
\caption{HV divider with maximum amplitude limitation. (a) MCP-PMT; (b) Dynode-PMT}\label{Figure 23}
\end{figure}

As shown in the previous sections, the maximum amplitude of the dynode-PMT is around 7-8\,V, but the maximum amplitude of the MCP-PMT is around 30-50\,V. In the early stage of JUNO electronics design, the maximum amplitude needed to be lower than the 8\,V required by the electronics interface. Therefore, we used a diode after the decoupling capacitor to limit the output amplitude and the maximum amplitude of the PMT, which was limited to 8\,V without affecting the signal bandwidth. The optimization of the HV divider with the scheme of limiting the maximum amplitude of the PMT has also been finalized, as shown in Fig.\,\ref{Figure 23}.

\subsection{Design II: Two outputs scheme}

\begin{figure}[!ht]
	\centering
	\includegraphics[width=6.0cm]{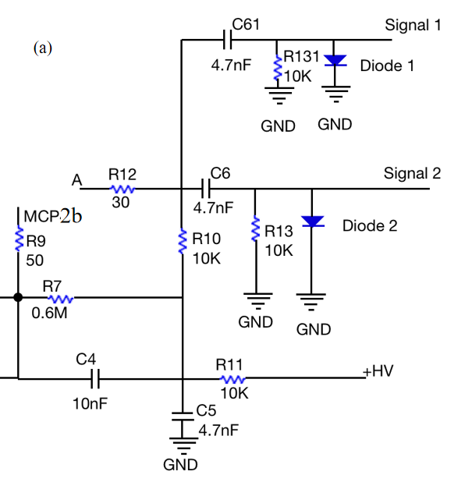}
   \includegraphics[width=6.0cm]{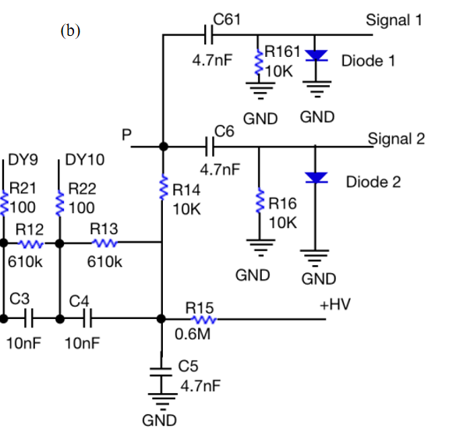}
	\caption{Anode readout with two outputs and maximum output amplitude limitation. (a) MCP-PMT; (b) Dynode-PMT}\label{Figure 24}
\end{figure}

During the R$\&$D of JUNO, another option was required to provide two identical outputs from the HV divider to the electronics \cite{JUNO-BX-BELLATO2021164600}. The two-output option, based on design I, was designed in the HV divider for overshoot suppression by the following electronics. The final design with the two-output scheme is shown in Fig.\,\ref{Figure 24}.

\subsection{Design III: BX2 scheme}
\begin{figure}[!htbp]
	\centering
	\includegraphics[width=11cm]{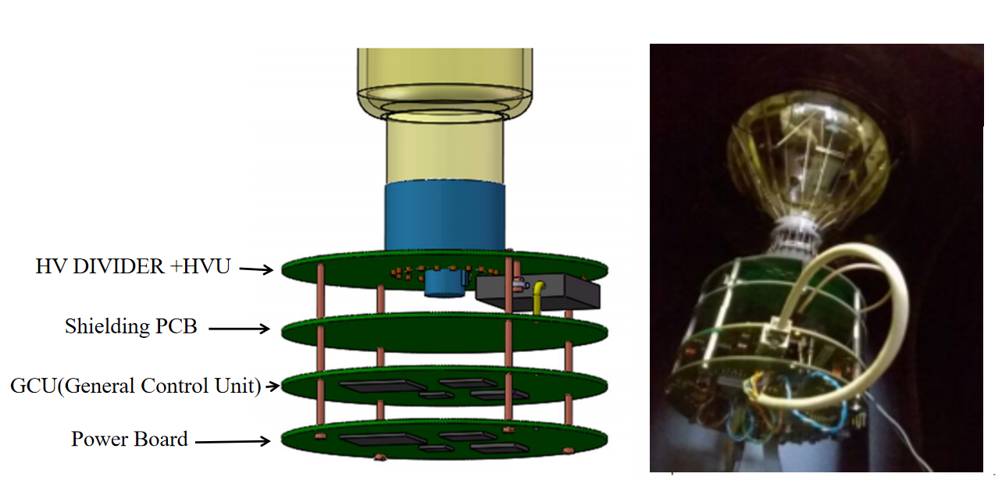}
	\caption{JUNO electronic pre-scheme of BX2 }\label{Figure 25}
\end{figure}

\begin{figure}[!ht]
	\centering
\includegraphics[width=13cm]{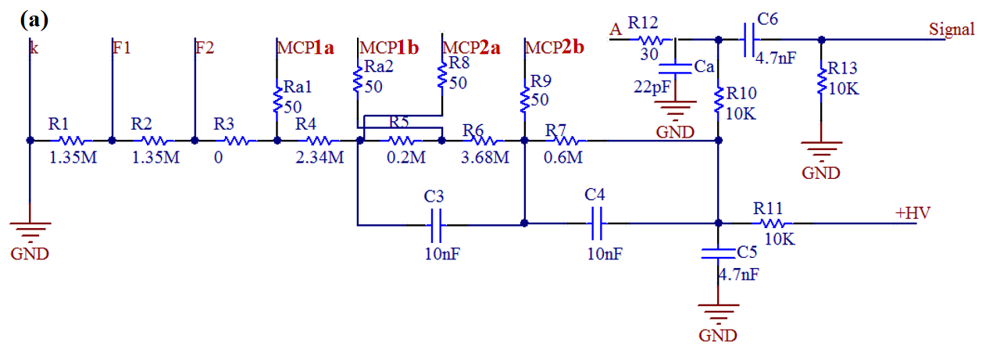}
 \includegraphics[width=13cm]{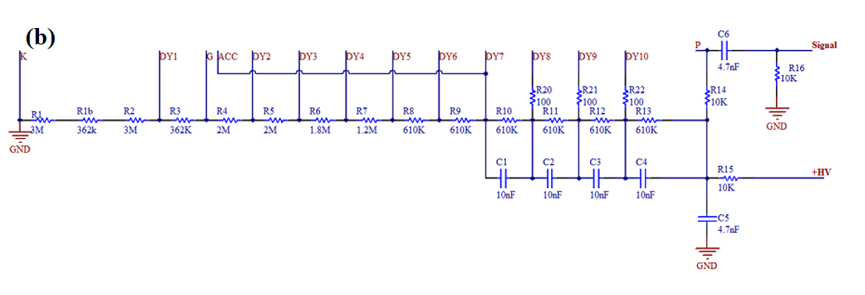}
	\caption{ HV divider of the BX2 scheme. (a) MCP-PMT; (b) Dynode-PMT}
     \label{fig:base:PMT:bx2}
\end{figure}

The BX2 scheme was another early option for the JUNO design. In the BX2 scheme, as shown in Fig.\,\ref{Figure 25}, the HV divider and the decoupling capacitor circuit (Splitter) of the PMT, as well as the high voltage and Front End Unit (FEU), are integrated. The final design of the HV divider with the BX2 scheme is shown in Fig.\,\ref{fig:base:PMT:bx2}, where it is only integrated with the decoupling capacitor and the amplitude limitation diode has been removed.With the BX scheme, there are many interfaces, such as the interface between the electronics and PMT waterproof potting, cables, etc. In addition, the power of each channel with the BX2 scheme is 17.5\,W, and there are a total of 20,000 channels. Furthermore, it needs to deal with a more complicated reliability problem.

\subsubsection{For PMT acceptance test}

\begin{figure}[!ht]
	\centering
	\includegraphics[width=10cm]{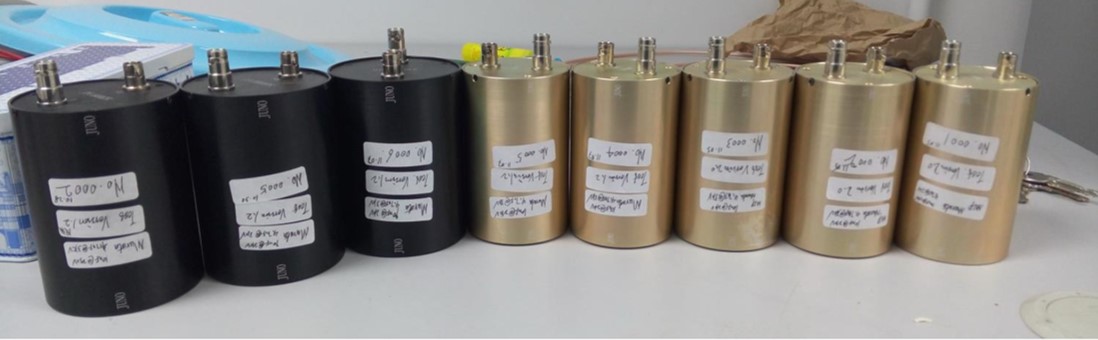}
	\caption{Plug-gable HV divider for bare PMT testing. Left for HPK PMTs; right for NNVT PMTs.}
     \label{fig:base:PMT:test}
\end{figure}

A special version of the PMT HV divider of the BX2 schema (only the HV divider as in Fig.\ref{fig:base:PMT:bx2}) with a decoupling capacitor of 10\,nF is required to support the acceptance testing of the bare PMTs, which is designed as a plug-gable version to test the bare PMTs \cite{JUNO:20inchPMT:2022hlz}.
The products used for bare PMT testing are shown in Fig.\,\ref{fig:base:PMT:test}.

 \subsection{Design IV: 1F3 scheme}

\begin{figure}[!htbp]
	\centering
	\includegraphics[width=11cm]{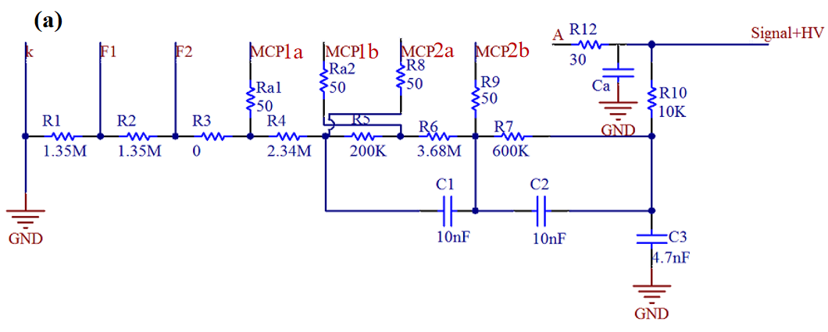}
 \includegraphics[width=11cm]{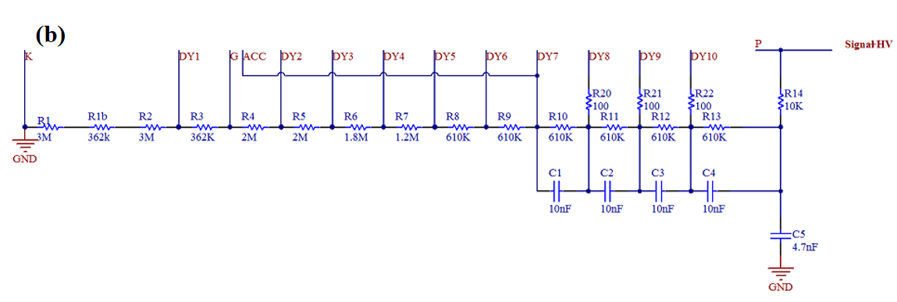}
	\caption{Final design of HV divider of 1F3 scheme. (a) MCP-PMT; (b) Dynode-PMT}\label{Figure 27}
\end{figure}

To avoid the disadvantages of the BX scheme and decouple the electronics and PMTs, a new scheme for the JUNO electronics system has been proposed, named 1F3 (one electronics box for three PMT channels) \cite{JUNO:physics:2022PrPNP.12303927J}.
In the 1F3 scheme, the HV divider of the PMT is separated from the splitter, FEC, and ADC, as compared to the BX scheme. The challenge of this design is the impedance matching between the PMT and the electronics \cite{Liu_2023_1F3}. The circuit for the 1F3 scheme is shown in Fig.\,\ref{Figure 27}.

\begin{figure}[!htbp]
	\centering
	\includegraphics[width=10cm]{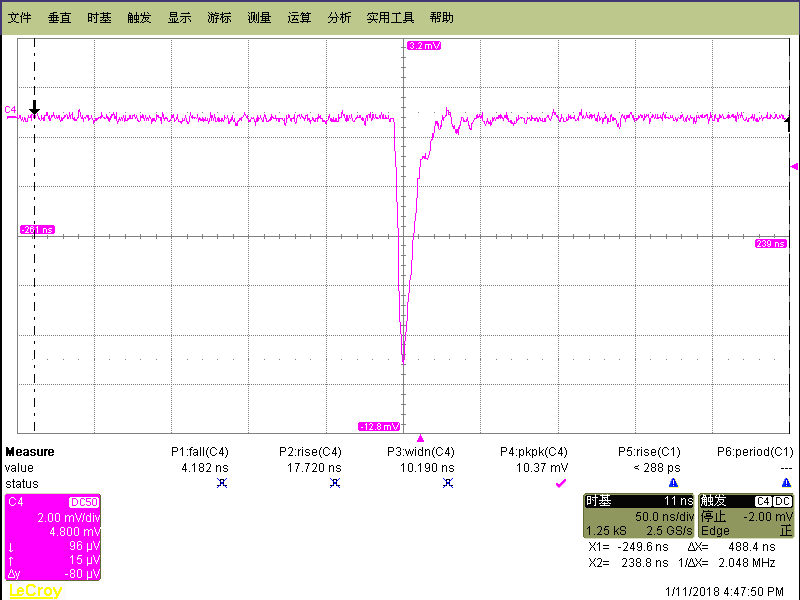}
	\caption{The waveform of SPE of MCP-PMT, R12=5\,$\Omega$}\label{Figure 28}
\end{figure}

In the 1F3 scheme, a 2\,m long high-voltage coaxial cable is used between the PMT and the electronic underwater box. The impedance of the cable is required to be 50\,$\Omega$, but the HV connector is not easy to match the requirement. When the impedance cannot match perfectly, the PMT output waveform will show some ringing, reflection, and a larger overshoot. According to the study of the MCP-PMT, R12 will affect the ringing.
When the R12 value is smaller, the ringing is larger, the SPE amplitude is larger, and the fall time is faster. R12=5\,$\Omega$ is recommended, and the ringing of the waveform of MCP-PMT is very small, as shown in Fig.\,\ref{Figure 28}.

\subsubsection{Final design for JUNO}

\begin{figure}[!htbp]
	\centering
	\includegraphics[width=13cm]{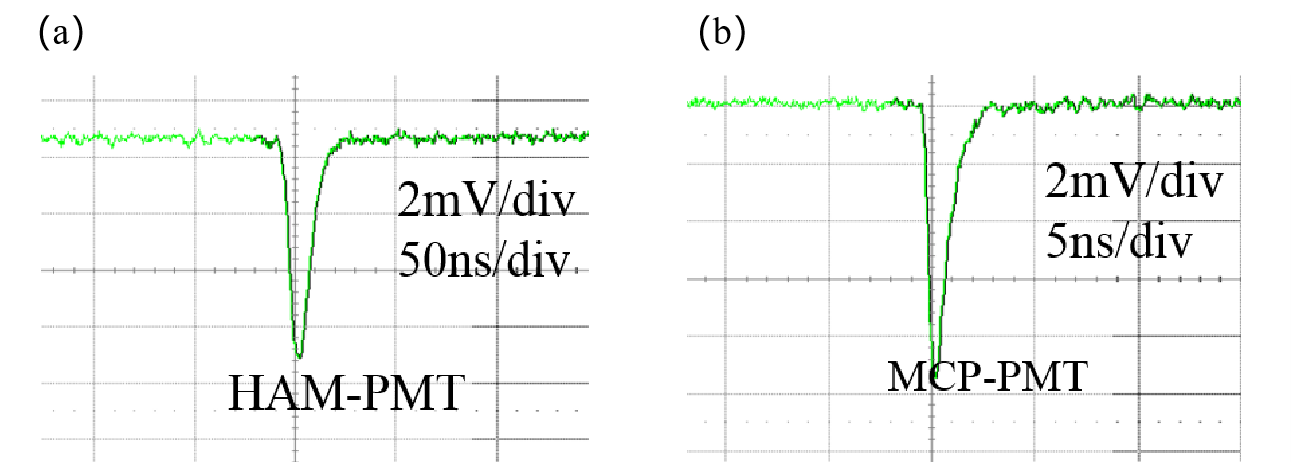}
	\caption{SPE waveforms. (a) Dynode-PMT; (b) MCP-PMT}\label{Figure 29}
\end{figure}

For JUNO, the 1F3 scheme was adopted to match the final electronics, and it was used as the final design of the HV dividers for the 20-inch PMT, as shown in Fig.\,\ref{Figure 27}. The SPE waveforms of dynode-PMT and MCP-PMT are shown in Fig.\,\ref{Figure 29}.

\begin{figure}[!htbp]
	\centering
	\includegraphics[width=4.8cm]{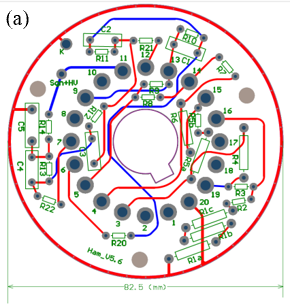}
 \includegraphics[width=5cm]{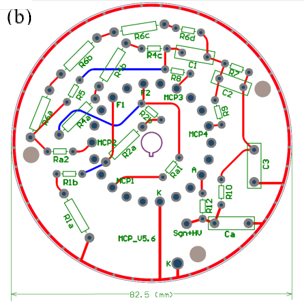}
	\caption{The PCB of JUNO PMT divider; (a) Dynode-PMT; (b) MCP-PMT}\label{Figure 30}
\end{figure}

In the JUNO, the PCB design of the HV divider with good reliability is also one of the key factors. Table \ref{table 8} shows the design requirements for the PCB. Fig.\,\ref{Figure 30} shows the PCB design for the JUNO PMT HV divider, where the resistors and capacitors on the PCB have been selected from Vishay's wire-wound components for reliability. The capacitors were selected after a pre-sample aging test from eight types of candidates.

\begin{table}  	
	\begin{center}
		\caption{The requirements of PCB design}\label{table 8}
		\resizebox{0.8\textwidth}{!}{
			\begin{tabular}{cc}
				\hline
				Width of PCB wring  &   0.5\,mm (Common wiring)\\
				&	0.7\,mm  \\
    				& (Wiring between  \\  
            				& high voltage and ground)  \\  
                \hline
				Through hole size &	Actual pin diameter+0.2\,mm \\ 
				Pad size &	2-2.5 times diameter of PMT Pin  \\ 
				Minimum distance design of breakdown/solder joint&	\textgreater1.6\,kV/mm 	(40\,V/mil) \\
                \hline            
		\end{tabular} }
	\end{center}
\end{table}

\section{Quality control \& burning test}
\label{sec:1:product}

We have finalized the design, optimization, and production of the JUNO HV divider. The performance acceptance testing of 20,000 20-inch PMTs has been finished in early 2021, and the final production of the HV divider for JUNO has also been completed.

\subsection{Reliability}

\begin{figure}[!ht]
    \includegraphics[width=6cm]{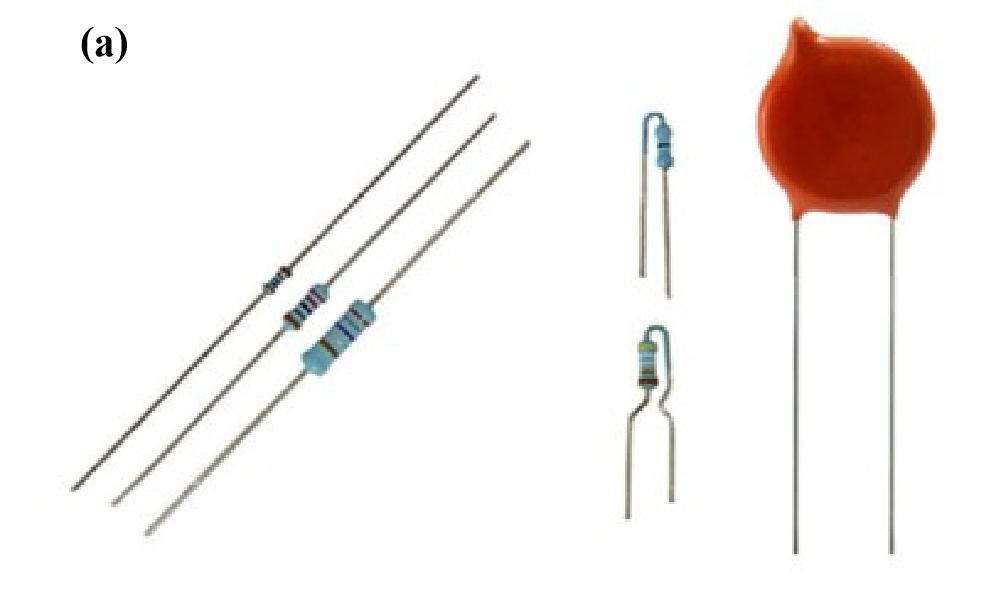}
    \includegraphics[width=7.5cm]{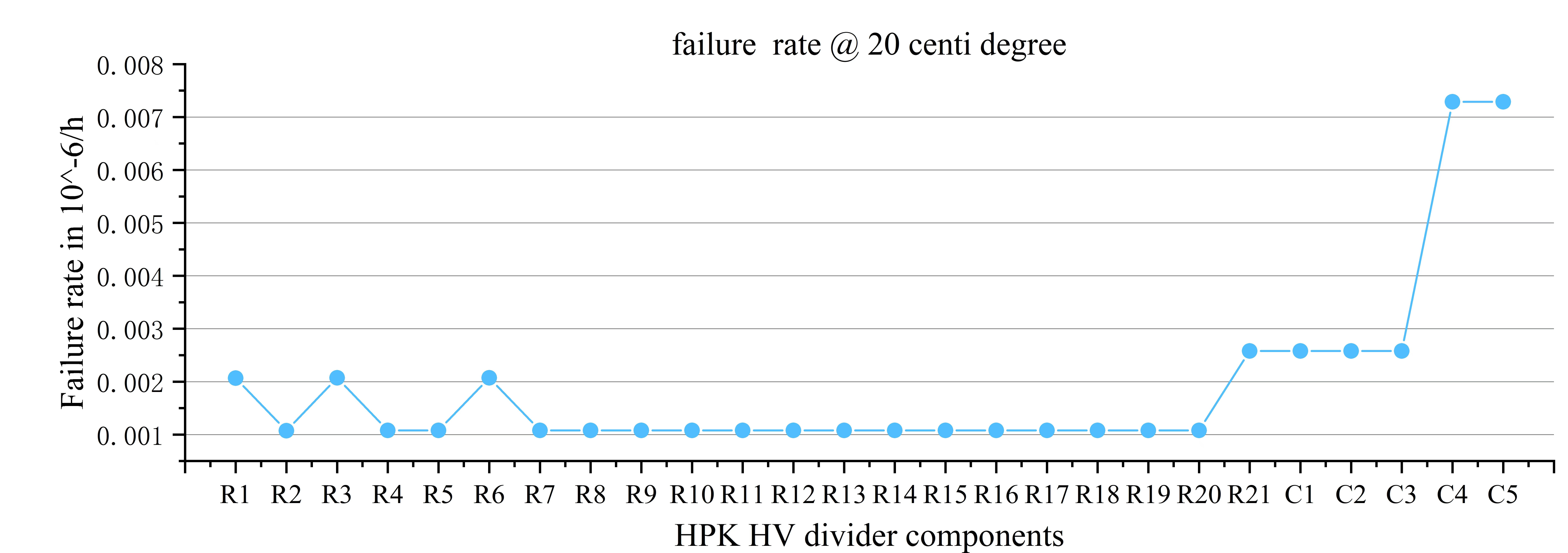}
    \includegraphics[width=7.5cm]{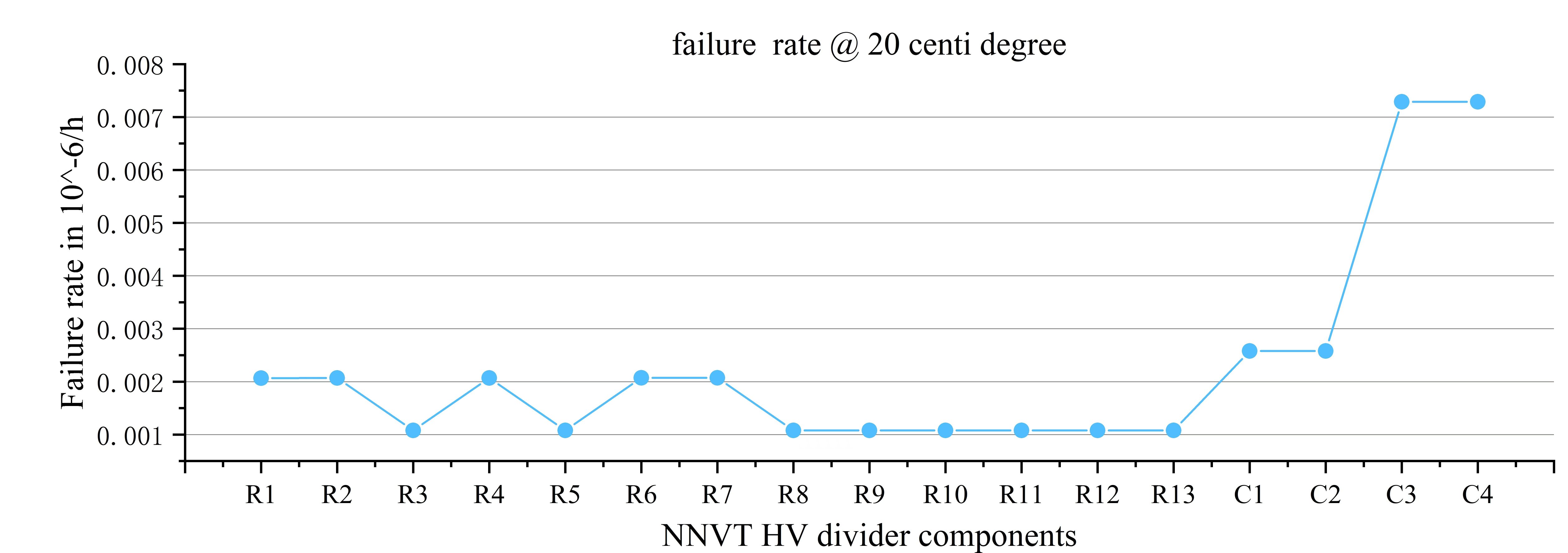}
	\caption{Components of HV divider and reliability calculation. (a) resistors and capacitors; (b) reliability calculation of each components for HPK dynode-PMT; (c) reliability calculation of each components for NNVT MCP-PMT.}
    \label{fig:divider:aging}
\end{figure}

For the JUNO, the HV divider is required to have a lifetime of over 20 years. With the selected components, a reliability expectation has been forecasted for the lifetime and possible failure rate of the HV dividers \cite{GJBZ-299C-2006}. This has been done with different component standards and temperatures, and the results with commercial components at 20$^{\circ}$C are shown in Fig.\,\ref{fig:divider:aging}. The capacitors contribute the maximum according to the model, even though the workload of each component in power and voltage is designed to be less than 1/2 of their nominal value. The final failure rate of each HV divider is less than 50\,FIT with commercial components, which already meets the requirements of JUNO.

\subsection{Soldering}

\begin{figure}[!ht]
	\centering
	\includegraphics[width=10cm]{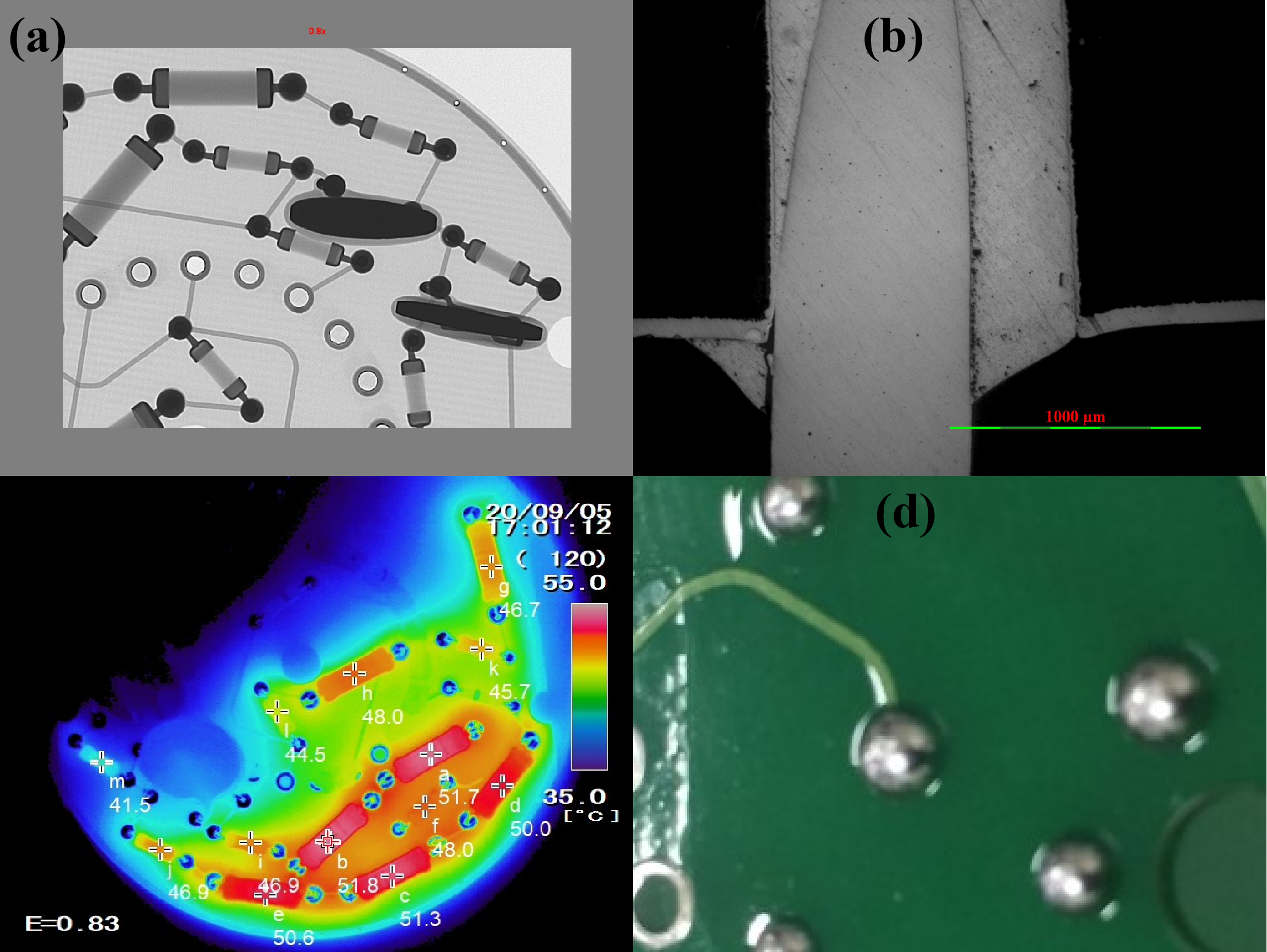}
\caption{HV divider soldering quality control. (a) X-ray check; (b) soldering quality; (c) temperature monitoring; (d) smooth soldering shape.}
    \label{fig:divider:soldering}
\end{figure}

In order to meet the requirements on the PMT flasher, special testing on the flasher of the HV divider has been conducted \cite{PMT-flasher-Yang_2020}. Smooth soldering and special cleanliness are required for the flasher, and the soldering quality is further checked with a special procedure and pre-samples, as shown in Fig.\,\ref{fig:divider:soldering}. The soldering, production, and following burning test of PCBs of HV dividers for 20,000 20-inch PMTs have been entrusted to Tianjin Centre Advanced Tech, CO, LTD \cite{tjcentre}. This ensures a high level of quality control and reliability for the production of the HV dividers. The soldering quality, temperature underworking, and maximum voltage of 6000\,V were applied to four random samples, and all of them were passed the test, as shown in Fig.\,\ref{fig:divider:soldering}. This indicates that the production process for the HV dividers is reliable and meets the design requirements.

\subsection{Burning test}

\begin{figure}[!ht]
    \includegraphics[width=6cm]{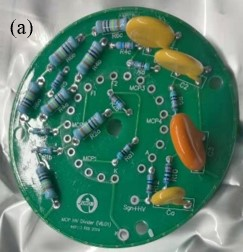}
    \includegraphics[width=8.2cm]{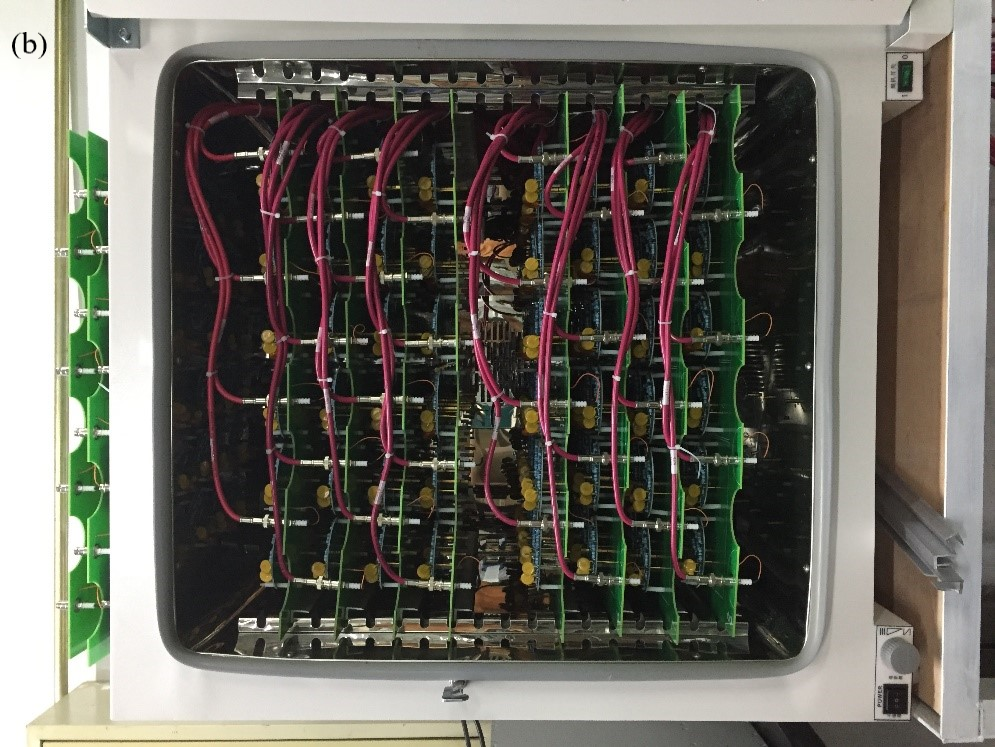}
	\includegraphics[width=12cm]{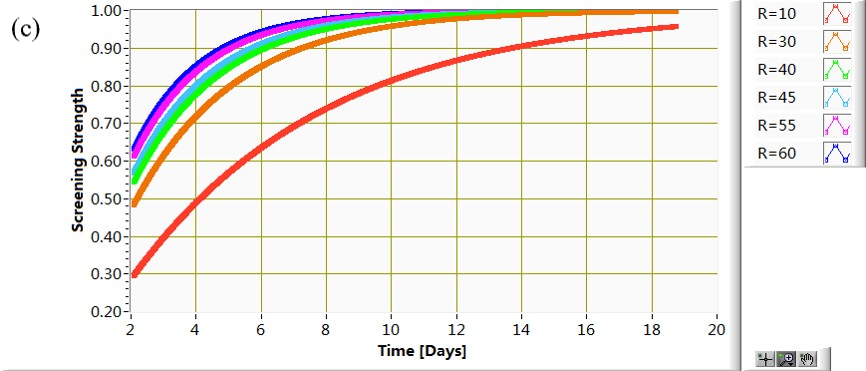}
	\caption{Burning test of the HV divider PCBs. (a) a sample of the produced divider; (b) The incubator and assembled samples of the burning test; (c) The burning strength calculation with different conditions;}
    \label{fig:divider:burning}
\end{figure}

After soldering and cleaning the PCBs, a burning test is carried out for each HV divider piece for 12 days $\times$ 24 hours at a high voltage workload of 3,000V and a temperature of 75$^{\circ}$C to reach 90\% strength, as shown in Fig.\,\ref{fig:divider:burning}. In total, 21,000 PCBs of HV dividers are subjected to aging experiments, and their aging performances are monitored by the current of the voltage divider. During the test, only four 362\,k$\Omega$ resistors of dynode-PMT were found to be broken, and the failure ratio was approximately 0.000067. The burning test effectively screens out defective components. The failed resistors were further checked by the vendor and another standalone analysis company, and the conclusion was that the failure was due to a possible random voltage surge. Even considering these kinds of failures, the system still meets the requirements of JUNO.

\section{Summary}
\label{sec:1:sum}

In summary, we have successfully designed, optimized, and produced the 20-inch PMT HV divider for the JUNO.  The HV divider is designed with low current and small overshoot and ringing for the HPK dynode-PMT. We also designed and developed the HV divider for the new large-area 20-inch MCP-PMT, with optimized HV ratio and DC to improve its collection efficiency, P/V, and charge resolution. The overshoot and ringing of the 20-inch MCP-PMTs have been effectively controlled, and the rise time and fall time have been improved to meet the JUNO requirements. Four schema designs of HV dividers for different options have been designed and optimized.The mass production of HV dividers for 20,000 JUNO 20-inch PMTs has been completed with good reliability after a burning test. The design, optimization, PCB design and production, soldering and aging test, and quality control of the HV divider have been presented in this article, which can provide valuable experiences for other photodetector experiments.

\appendix

\acknowledgments

This work was supported by the National Natural Science Foundation (NSFC) of China No. 11875282, the Strategic Priority Research Program of the Chinese Academy of Sciences, Grant No.\,XDA10011100, the CAS Center for Excellence in Particle Physics and  the Research Foundation of Education Bureau of Hunan Province, China (Grant No. 22A0286).










\bibliographystyle{unsrt}
\bibliography{allcites}   

\end{document}